\documentclass[cmp,final]{svjour}
\usepackage{amsmath}
\usepackage{amsfonts}
\usepackage{amssymb}
\usepackage{psfig}

\journalname{Communications in Mathematical Physics}

\spnewtheorem*{hypothesis}{Hypotheses}{\bf}{\it}

\begin{document}
\title{Reconstructing the Thermal Green Functions at Real Times from Those at Imaginary Times}

\titlerunning{Reconstructing Thermal Green Functions}
\authorrunning{G. Cuniberti, E. De Micheli, G. A. Viano}

\author{Gianaurelio Cuniberti\inst{1}, Enrico De Micheli\inst{2}, Giovanni Alberto Viano\inst{3}}
\institute{Max--Planck--Institut f\"ur Physik komplexer Systeme,
N\"othnitzer Stra{\ss}e 38, 01187 Dresden, Germany. \\
\email{cunibert@mpipks-dresden.mpg.de}
\and
Istituto di Cibernetica e Biofisica - Consiglio Nazionale delle Ricerche, 
Via De Marini 6, 16149 Genova, Italy. \\
\email{demic@icb.ge.cnr.it}
\and
Dipartimento di Fisica - Universit\`a di Genova and Istituto Nazionale di Fisica Nucleare \\
Via Dodecaneso, 33 - 16146 Genova, Italy.
\email{viano@ge.infn.it}
}

\date{Received: 17 February 2000 / Accepted: 12 July 2000
\\ published in Communications in Mathematical Physics \textbf{216}, 59--83 (2001)}
\communicated{D.C. Brydges}

\maketitle

\newfont{\eur}{eurm10}

 \newcommand{\lab} {\left\langle}
 \newcommand{\rab} {\right\rangle}
\newcommand{\R}{{\mathbb R}}
\newcommand{\C}{{{\mathbb C}\,}}
\newcommand{\Z}{{\mathbb Z}}
\newcommand{\N}{{\mathbb N}}
\newcommand{\HH}{{{\mathbb H}\,}}

\newcommand{\eurA}{\mbox{\eur A}}
\newcommand{\eurR}{\mbox{\eur R}}

\newcommand{\cG}{{\cal G}}
\newcommand{\cI}{{\cal I}}

\newcommand{\tJ}{\tilde{J}}
\newcommand{\tcG}{\tilde{\cal G}}

\newcommand{\beq}{\begin{eqnarray}}
\newcommand{\eeq}{\end{eqnarray}}

\newcommand{\Imag}{{\mbox{Im}\,}}
\newcommand{\Real}{{\mbox{Re}\,}}
\newcommand{\Trace}{{\mbox{Tr}\,}}

\newcommand{\UU}{{\bf 1}}

\newcommand{\Cir}[1]{\stackrel{\scriptscriptstyle\hspace{.2ex}\circ}{#1}}
\newcommand\staccrel[2]{\mathrel{\mathop{#1}\limits_{#2}}}

\begin{abstract}
By exploiting the analyticity and boundary value properties of the thermal Green 
functions that result from the KMS condition in both time and energy complex 
variables, we treat the general (non--perturbative) problem of recovering the 
thermal functions at real times from the corresponding functions at imaginary times,
introduced as primary objects in the Matsubara formalism. The key property on 
which we rely is the fact that the Fourier transforms of the retarded and advanced 
functions in the energy variable have to be the ``unique Carlsonian analytic 
interpolations'' of the Fourier coefficients of the imaginary--time correlator, 
the latter being taken at the discrete Matsubara imaginary energies, respectively 
in the upper and lower half--planes. Starting from the Fourier coefficients 
regarded as ``data set'', we then develop a method based on the Pollaczek 
polynomials for constructing explicitly their analytic interpolations.
\end{abstract}

\section{Introduction}
\label{se:introduction}
In the standard imaginary--time formalism of quantum statistical mechanics (tracing 
back to Matsubara~\cite{Matsubara}) and, later on, of quantum field theory at 
finite temperature (see e.g.~\cite{LeBellac} and references therein), there arises 
the a--priori non--trivial problem of recovering the ``physical'' correlations at 
real times starting from data at imaginary times. More specifically, the correlations 
at imaginary--time of observables (or, more generally, of boson or fermion fields) 
in a thermal equilibrium state at temperature $T=\beta^{-1}$ are defined as periodic 
(or antiperiodic) functions of period $\beta$, and therefore they are equivalently 
characterized by their {\it discrete} mode expansion 
$\frac{1}{\beta}\sum_n \cG_n\exp(-i\zeta_n\tau)$ in terms of the so--called 
``Matsubara energies'' $i\zeta_n$, where $\zeta_n=2n\pi/\beta$ (or $(2n+1)\pi/\beta$).

The problem of recovering the correlations at real time, or equivalently the 
retarded and advanced Green's functions at real energies, from the previous 
sequence of Fourier coefficients $\{\cG_n\}$ admits a unique and well--defined 
theoretical solution in terms of the notion of ``Carlsonian analytic interpolation 
of this sequence''. This can be achieved~\cite{Bros1}, and will be recalled below 
in Sect.~\ref{se:double}, if the imaginary--time formalism is embedded in the 
(conceptually more satisfactory) general description of quantum thermal states as KMS states~\cite{Haag1}. However, as suggested by the lattice approach of the imaginary--time formalism, it may be interesting to have a {\it concrete} procedure 
for constructing satisfactory approximate solutions of this problem when one 
starts from incomplete data sets.

In this paper we give a precise algorithm for the previous reconstruction problem; 
this mathematical method is presented in Sect.~\ref{se:representation}. 
Moreover, in the subsequent Sect.~\ref{se:reconstruction}, the method is applied 
to the case when the data are finite in number and affected by noise.

Let us consider the algebra ${\cal A}$ generated by the observables of a quantum 
system. Denoting by $A,B,\ldots$ arbitrary elements of ${\cal A}$ and by 
$A\rightarrow A(t)$ ($A=A(0)$) the action of the (time--evolution) group of 
automorphisms on this algebra, we now recall the KMS analytic structure of 
two--point correlation functions $\lab A(t_1) B(t_2) \rab_{\Omega_\beta}$, in a thermal 
equilibrium state $\Omega_\beta$ of the system at temperature $T=\beta^{-1}$.

By time--translation invariance, these quantities only depend on $t=t_1-t_2$, 
and we shall put
\begin{eqnarray}
\label{intro_1}
{\cal W}_{AB}(t) = \lab A(t)\,B \rab _{\Omega_\beta}, \\
{\cal W}'_{AB}(t) = \lab B\,A(t) \rab _{\Omega_\beta}.
\end{eqnarray}
In finite volume approximations, the time--evolution is represented by a unitary 
group $e^{iH't}$, so that
\begin{eqnarray}
\label{intro_2}
A(t) = e^{iH't} A \, e^{-iH't},
\end{eqnarray}
where $H'=H-\mu N$, $H$ being the Hamiltonian, $\mu$ the chemical potential, 
and $N$ the particle number; under general conditions, the operators $e^{-\beta H'}$ 
have finite traces for all $\beta>0$ (see e.g.~\cite{Haag1}). Then the correlation 
functions are given, correspondingly, by the formulae
\begin{eqnarray}
\label{intro_3}
{\cal W}_{AB}(t) = \frac{1}{Z_\beta}\Trace\left\{e^{-\beta H'}A(t)\,B\right\},  \\
{\cal W}'_{AB}(t) = \frac{1}{Z_\beta}\Trace\left\{e^{-\beta H'} B A(t)\right\},
\end{eqnarray}
where $ Z_\beta=\Trace e^{-\beta H'} $.

One then introduces the following holomorphic functions of the complex time 
variable $t+i\gamma$:
\beq
\label{intro_4}
G_{AB}(t+i\gamma)= \frac{1}{Z_\beta}\Trace\left\{e^{-(\beta+\gamma) H'} 
A(t)\,e^{\gamma H'} B\right\},
\eeq
analytic in the strip $\{t+i\gamma;\,t\in\R,\,-\beta<\gamma<0\}$, and
\beq
\label{intro_5}
G'_{AB}(t+i\gamma)= \frac{1}{Z_\beta}\Trace\left\{e^{-(\beta-\gamma) H'} 
B\,e^{-\gamma H'} A(t)\right\},
\eeq
analytic in the strip $\{t+i\gamma;\,t\in\R,\,0<\gamma<\beta\}$, which are such that:
\begin{eqnarray}
\label{intro_6}
\lim_{\staccrel{\scriptstyle \gamma\rightarrow 0}{\scriptstyle \gamma < 0}} 
G_{AB}(t+i\gamma)={\cal W}_{AB}(t), \\
\lim_{\staccrel{\scriptstyle \gamma\rightarrow 0}{\scriptstyle \gamma > 0}} 
G'_{AB}(t+i\gamma)={\cal W}'_{AB}(t).
\end{eqnarray}
From (\ref{intro_4}), (\ref{intro_5}) and the cyclic property of $\Trace$, we then 
obtain the KMS relation
\beq
\label{intro_7}
{\cal W}_{AB}(t)=\Trace e^{-\beta H'} A(t)\,B = \Trace B\, e^{-\beta H'} A(t)
= G'_{AB}(t+i\beta),
\eeq
which implies the identity of holomorphic functions (in the strip $0<\gamma<\beta$)
\beq
\label{intro_8}
G_{AB}(t+i(\gamma-\beta)) = G'_{AB}(t+i\gamma).
\eeq
According to the analysis of~\cite{Haag1} in the Quantum Mechanical framework and of 
\cite{Buchholz} in the Field--theoretical framework, this KMS analytic structure is 
preserved by the thermodynamic limit under rather general conditions.

In the case when the algebra ${\cal A}$ is generated by smeared--out bosonic or 
fermionic field operators (field theory at finite temperature), the principle of 
relativistic causality of the theory implies additional relations for the corresponding 
pairs of analytic functions $(G,G')$. In fact, this principle of relativistic causality 
is expressed by the {\it commutativity} (resp. {\it anticommutativity}) {\it relations}
for the boson field $\Phi({\bf x})$ (resp. fermion field $\Psi({\bf x})$) at 
space--like separation:
\beq
\label{intro_9}
\left[\Phi(t,\vec{x}),\Phi(t',\vec{x'})\right]=0~~~~~
({\rm resp.}~\left\{\Psi(t,\vec{x}),\Psi(t',\vec{x'})\right\}=0)~~~{\rm for}~~ 
(t-t')^2<(\vec{x}-\vec{x'})^2.
\eeq
In this field--theoretical case, we can choose as suitable operators $A$ the 
``smeared--out field operators'' of the form 
$A=\int\Phi(y_0,\vec{y})f(y_0,\vec{y})dy_0\,d\vec{y}$ 
(resp. $\int\Psi(y_0,\vec{y})f(y_0,\vec{y})dy_0\,d\vec{y}$), where $f$ is any 
smooth test--function with (arbitrary small) compact support around the origin 
in space--time variables. For the observable $B$, we can then choose any operator 
$A_{\vec{x}}$ obtained from $A$ by the action of the space--translation group 
(which amounts to replace the test--function $f(y_0,\vec{y})$ by
$f(y_0,\vec{y})=f(y_0,\vec{y}-\vec{x})$). It then follows from (\ref{intro_9}) 
that the corresponding analytic functions $G_{AA_{\vec{x}}}(t+i\gamma)$ and 
$G'_{AA_{\vec{x}}}(t+i\gamma)$ (satisfying (\ref{intro_8})) have real boundary 
values ${\cal W}_{AA_{\vec{x}}}(t)$ and ${\cal W}'_{AA_{\vec{x}}}(t)$ which satisfy,
on some interval $|t|<t(\vec{x},f)$, {\it coincidence relations} of the following form:
\begin{eqnarray}
\label{intro_10}
{\cal W}_{AA_{\vec{x}}}(t)& = &{\cal W}'_{AA_{\vec{x}}}(t)~~~~~~
\mbox{in the boson case,} \\
\label{intro_11}
{\cal W}_{AA_{\vec{x}}}(t)& =& -{\cal W}'_{AA_{\vec{x}}}(t)~~~~~~
\mbox{in the fermion case.}
\end{eqnarray}
Then, in view of identity (\ref{intro_8}), the coincidence relations (\ref{intro_10}) 
and (\ref{intro_11}) imply the existence of a single analytic function 
$\cG_{AA_{\vec{x}}}(t+i\gamma)$ which is such that:
\begin{itemize}
\item[a)] in the boson case:
\begin{eqnarray}
\label{intro_12}
\cG_{AA_{\vec{x}}} & = & G_{AA_{\vec{x}}}~~~~~
\mbox{for}~-\beta < \gamma < 0, \\
\label{intro_13}
\cG_{AA_{\vec{x}}} & = & G'_{AA_{\vec{x}}}~~~~~
\mbox{for}~0 < \gamma < \beta;
\end{eqnarray}
\item[b)] in the fermion case:
\begin{eqnarray}
\label{intro_14}
\cG_{AA_{\vec{x}}} & = & G_{AA_{\vec{x}}}~~~~~\mbox{for}~-\beta < \gamma < 0, \\
\label{intro_15}
\cG_{AA_{\vec{x}}} & = & -G'_{AA_{\vec{x}}}~~~~~\mbox{for}~0 < \gamma < \beta.
\end{eqnarray}
\end{itemize}
Correspondingly, it follows that $\cG_{AA_{\vec{x}}}$ is either periodic or 
antiperiodic with period $i\beta$ in the full complex plane minus periodic cuts 
along the half--lines $\{t+i\gamma;\,t>t(\vec{x},f),\,\gamma=k\beta,\,k\in\Z\}$ and 
$\{t+i\gamma;\,t<-t(\vec{x},f),\,\gamma=k\beta,\,k\in\Z\}$.

These analytic functions $\cG_{AA_{\vec{x}}}(t+i\gamma)$ are smeared--out forms 
(corresponding to various test--functions $f$) of the thermal two--point function 
of the fields $\Phi$ (or $\Psi$) in the complex time variable. In other words, 
this thermal two--point function can be fully characterized in terms of an analytic
function ${\cal G}(t+i\gamma,\vec{x})$ (with regular dependence in the space variables) 
enjoying the following properties:
\begin{itemize}
\item[a)] ${\cal G}(t+i\gamma,\vec{x}) = \epsilon \, {\cal G}(t+i(\gamma-\beta),\vec{x})$, 
where $\epsilon=+$ for a boson field, and $\epsilon=-$ for a fermion field;
\item[b)] for each $\vec{x}$, the domain of ${\cal G}$ in the complex variable $t$ is
$\C \setminus \{t+i\gamma;\,|t|>|\vec{x}|;\,\gamma=k\beta,\,k\in\Z\}$;
\item[c)] the boundary values of ${\cal G}$ at real times are the thermal correlations 
of the field, namely:
\begin{eqnarray}
\label{intro_16}
\lim_{\staccrel{\scriptstyle \gamma\rightarrow 0}{\scriptstyle \gamma < 0}}
{\cal G}(t+i\gamma,\vec{x})& =& {\cal W}(t,\vec{x}), \\
\label{intro_17}
\lim_{\staccrel{\scriptstyle \gamma\rightarrow 0}{\scriptstyle \gamma > 0}}
{\cal G}(t+i\gamma,\vec{x})& =& {\cal W}'(t,\vec{x}),
\end{eqnarray}
where in finite volume regions, ${\cal W}$ and ${\cal W}'$ can be formally expressed 
as follows (a rigorous justification of the trace--operator formalism in the 
appropriate Hilbert space being given in~\cite{Buchholz}):
\begin{eqnarray}
\label{intro_18}
{\cal W}(t,\vec{x})=\frac{1}{Z_\beta}\,\Trace e^{-\beta H}\, \Phi(t,\vec{x})\, \Phi(0,\vec{0}), \\
\label{intro_19}
{\cal W}'(t,\vec{x})=\frac{1}{Z_\beta}\,\Trace e^{-\beta H}\, \Phi(0,\vec{0})\, \Phi(t,\vec{x}),
\end{eqnarray}
for the boson case, and similarly in terms of $\Psi(t,\vec{x})$ for the fermion case.
\end{itemize}
In this analytic structure, we shall distinguish two quantities that play an important role:
\begin{itemize}
\item[i)] the restriction ${\cal G}(i\gamma,\vec{x})$ of the function ${\cal G}$ to 
the imaginary axis is a $\beta$--periodic (or antiperiodic) function of $\gamma$ 
which must be identified with the ``{\it time--ordered product at imaginary times}'', 
considered in the Matsubara approach of imaginary--time formalism. In the latter, 
this quantity or its set of Fourier coefficients plays the role of {\it initial data}.
\item[ii)] The ``retarded'' and ``advanced'' two--point functions
\begin{eqnarray}
\label{intro_20}
{\eurR}(t,\vec{x})&=&i\,\theta(t)[{\cal W}(t,\vec{x})-\epsilon\, {\cal W}'(t,\vec{x})], \\
\label{intro_21}
{\eurA}(t,\vec{x})&=&-i\,\theta(-t)[{\cal W}(t,\vec{x})-\epsilon\, {\cal W}'(t,\vec{x})],
\end{eqnarray}
which are respectively the ``jumps'' of the function ${\cal G}$ across the real cuts
$\{t;\,t\geq |\vec{x}|\}$ and $\{t;\,t < -|\vec{x}|\}$. These kernels have an 
important causal interpretation; in particular, $\eurR$ describes the 
``response of the system'' to small perturbations of the equilibrium state. 
The knowledge of $\eurR$ and $\eurA$ and, consequently, of 
${\cal W}-{\cal W}'= -i\,(\eurR-\eurA)$ allows one to reconstruct ${\cal W}$ and 
${\cal W}'$ by the application of the Bose--Einstein factor $1/(1-e^{\mp\beta\omega})$ 
to their Fourier transforms $\widetilde{{\cal W}}(\omega)$, 
$\widetilde{{\cal W}'}(\omega)$ (this procedure being an implementation of the KMS 
property in the energy variable $\omega$).
\end{itemize}
The rest of the paper is devoted to the problem of {\it recovering the 
``real--time quantities'' $\eurR$ and $\eurA$, starting from the ``time--ordered 
product at imaginary times'' as initial data}. This will require the conjoint use 
of the analytic structure of ${\cal G}$ in complex time and of its Fourier--Laplace 
transform in the complex energy variable. In fact, the key property on which our 
reconstruction of real--time quantities relies is the following one: the 
Fourier--Laplace transforms $\widetilde{\eurR}$ and $\widetilde{\eurA}$ of the 
functions $\eurR$ and $\eurA$, which are defined and analytic respectively in the 
upper and lower half--planes of the energy variable $\omega$, are {\it analytic 
interpolations of the set of Fourier coefficients $\{{\cal G}_n\}$} of the function 
${\cal G}$ at imaginary times, the latter being taken at the Matsubara energies 
$\omega=i\,\zeta_n$. Moreover, the uniqueness of this interpolation is ensured by 
global bounds on $\widetilde{\eurR}$ and $\widetilde{\eurA}$, according to a standard 
theorem by Carlson~\cite{Boas}. The basic equalities that relate 
$\widetilde{\eurR}(i\,\zeta_n)$ and $\widetilde{\eurA}(i\,\zeta_n)$ to the 
corresponding coefficients ${\cal G}_n$ will be called ``Froissart--Gribov--type equalities'' 
for the following historical reason.
A general $n$--dimensional mathematical study of the type of double--analytic structure 
encountered here has been performed in~\cite{Bros2} in connection with the theory of 
complex angular momentum, where the original Froissart--Gribov equalities had been 
first discovered (in the old framework of S-matrix theory). The fact that this 
structure is relevant (in its simplest one--dimensional form) in the analysis of 
thermal quantum states has been already presented in~\cite{Bros1} in the framework 
of Quantum Field Theory at finite temperature.

\section{Double Analytic Structure of the Thermal Green Function and Froissart--Gribov--type Equalities}
\label{se:double}
In the following mathematical study we replace the complex time variable $t+i\gamma$ 
of the introduction by $\tau=i(t+i\gamma)$ in such a way that, in our 
``reconstruction problem'' treated in Sects.~\ref{se:representation} and 
\ref{se:reconstruction}, the {\it initial data} of the function ${\cal G}(\tau,\cdot)$ 
considered below correspond to {\it real values of $\tau$}. Up to this change of 
notation, this general analytic function ${\cal G}(\tau,\cdot)$ can play the role of 
the previously described two--point function of a boson or fermion field at fixed 
$\vec{x}$. However, since the only variables involved in the forthcoming study are 
$\tau$ and its Fourier--conjugate variable $\zeta$, the extra ``spectator variables'', 
denoted by the point $(\cdot)$, may as well represent a fixed momentum (after Fourier 
transformation with respect to the space variables) or the action on a test--function 
$f$ (as for the correlations of field observables $A=A(f)$ described in the introduction).

\begin{figure}[t]
\begin{center}
\centerline{\psfig{file=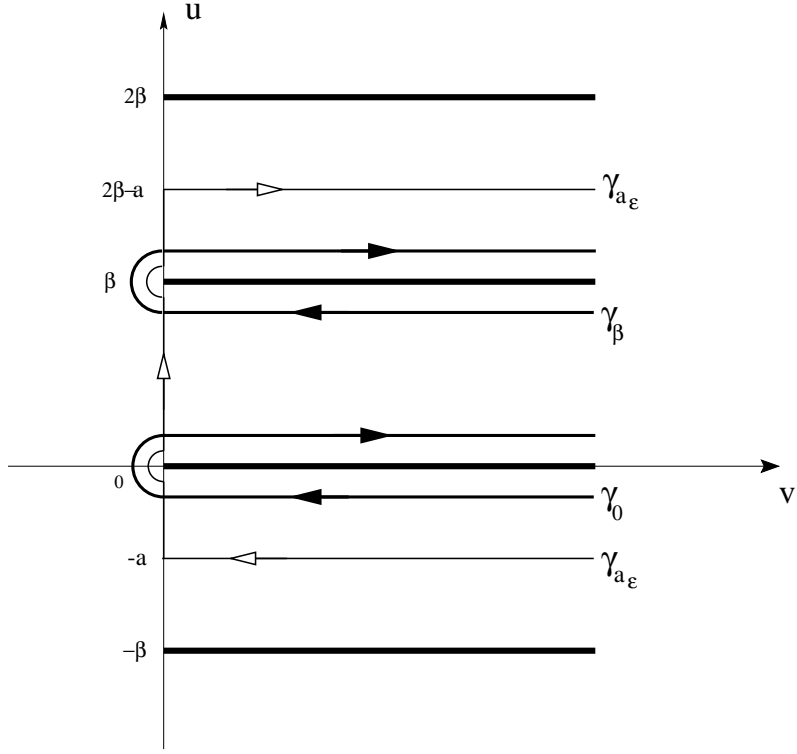,width=.70 \linewidth}}
\caption{Integration paths used in the proof of Theorem~\ref{theorem1}}
\label{figura_1}
\end{center}
\end{figure}

Let us summarize the analytic structure that we want to study.
\begin{hypothesis}
\label{hypo}
The function $\cG(\tau,\cdot)$, $(\tau=u+iv,\,u,v\in\R)$, satisfies the
following properties:
\begin{itemize}
\item[a{\rm)}] it is analytic in the open strips $k\beta<u<(k+1)\beta$ 
$(v\in\R,\,k\in\Z,\,\beta=1/T)$ and continuous at the boundaries;
\item[b{\rm)}] it is periodic {\rm(}antiperiodic{\rm)} for bosons {\rm(}fermions{\rm)} 
with period $\beta$, i.e.
\begin{equation}
\label{nove}
\cG(\tau+\beta,\cdot) = \left\{
\begin{array}{ll}
\cG(\tau,\cdot) & ~\mbox{for bosons,}~~~~~ (\tau\in\C), \\
-\cG(\tau,\cdot) & ~\mbox{for fermions,}~~~~~ (\tau\in\C);  \\
\end{array}
\right .
\nonumber
\end{equation}
\end{itemize}
\begin{eqnarray}
\label{dodici}
{\it c{\rm)}} & \hspace{1.7cm} &
\sup_{-k\beta<u<(k+1)\beta} \left | \cG(u+iv,\cdot)\right | \leq C |v|^\alpha,~~~~~~ 
(v\in\R;\,C,\alpha~\mbox{constants}).\hspace{2.4truecm}
\end{eqnarray}
\end{hypothesis}
We shall treat both the boson and fermion field cases at the same time by exploiting 
the $2\beta$--periodicity of the function $\cG(\tau,\cdot)$. To this purpose, we 
take the Fourier series (in the sense of $L^2[-\beta,\beta]$) of $\cG(\tau,\cdot)$, 
which we write
\beq
\label{tredici}
\cG(\tau,\cdot)=\frac{1}{2\beta}\sum_{n=-\infty}^{+\infty}\cG_n(\cdot) e^{-i\zeta_n\tau},
~~~\zeta_n=\frac{\pi}{\beta}\,n\,,
\eeq
and whose Fourier coefficients are given by
\beq
\label{quattordici}
\cG_n(\cdot) = \int_{-\beta}^\beta \cG(\tau,\cdot) \, e^{i\zeta_n\tau}\, d\tau.
\eeq
It is convenient to split expansion (\ref{tredici}) into two terms as follows:
\begin{eqnarray}
\cG^{(+)}(\tau,\cdot) & = & \frac{1}{2\beta}\sum_{n=0}^{+\infty}\cG^{(+)}_n(\cdot)\, 
e^{-i\zeta_n\tau},~~~ \cG^{(+)}_n(\cdot) \equiv \cG_n(\cdot),~~(n=0,1,2,\ldots), 
\label{quindicia} \\
\cG^{(-)}(\tau,\cdot) & = & \frac{1}{2\beta}\sum_{n=-1}^{-\infty}\cG^{(-)}_n(\cdot)\, 
e^{-i\zeta_n\tau},~~~ \cG^{(-)}_n(\cdot) \equiv \cG_n(\cdot),~~(n=-1,-2,\ldots), 
\label{quindicib}
\end{eqnarray}
then,
\begin{eqnarray}
\cG^{(+)}_n(\cdot) & = & \int_{-\beta}^\beta \cG^{(+)}(\tau,\cdot)\, e^{i\zeta_n\tau}\, d\tau,
~~~~~(n=0,1,2,\ldots),\label{newsedicia} \\
\cG^{(-)}_n(\cdot) & = & \int_{-\beta}^\beta \cG^{(-)}(\tau,\cdot)\, e^{i\zeta_n\tau}\, d\tau,
~~~~~(n=-1,-2,\ldots). \label{newsedicib}
\end{eqnarray}
We now introduce in the complex plane of the variable $\tau=u+iv$ $(u,v \in\R)$ 
the following domains: the half--planes
$\cI_\pm = \{\tau\in\C\,|\,\Imag\tau\gtrless 0\}$; the ``cut--domain'' $\cI_+\setminus\Xi_+$,
where the cuts $\Xi_+$ are given by $\Xi_+=\{\tau\in\C\,|\,\tau=k\beta+iv,\,v\geq 0,\,k\in\Z\}$,
and $\cI_-\setminus\Xi_-$, where $\Xi_-=\{\tau\in\C\,|\,\tau=k\beta+iv,\,v\leq 0,\,k\in\Z\}$.
Moreover, we denote by $\Cir{{\hspace{-0.5ex}A}}$ any subset $A$ of $\C$ which is 
invariant under the translation by $k\beta$, $k\in\Z$ (e.g. $\Cir{\Xi}_\pm$, 
$\cI_\pm\setminus \Cir{\Xi}_\pm$, etc.) (see Ref.~\cite{Bros2}I). Accordingly, 
the periodic cut--$\tau$--plane $\C\setminus (\Cir{\Xi}_+ \cup \Cir{\Xi}_-)$ will 
be denoted by $\Cir{\Pi}_\tau$. We now introduce the jump functions 
$J_{(k\beta)}^{(+)}(v,\cdot)$ and $J_{(k\beta)}^{(-)}(v,\cdot)$ that represent 
the discontinuities of $\cG^{(+)}(\tau,\cdot)$ and $\cG^{(-)}(\tau,\cdot)$ across 
the cuts located respectively at $\Real\tau\equiv u = k\beta$, $v\geq 0$, and at 
$\Real\tau\equiv u = k\beta$, $v\leq 0$, ($k\in\Z$):
\begin{eqnarray}
J_{(k\beta)}^{(+)}(v,\cdot) & = & 
+i \lim_{\staccrel{\scriptstyle \epsilon\rightarrow 0}{\scriptstyle \epsilon>0}}
\left\{\cG^{(+)}(k\beta+\epsilon+iv,\cdot)- \cG^{(+)}(k\beta-\epsilon+iv,\cdot) \right\},~~~~
(v\geq 0,\; k\in\Z), \label{diciassettea} \\
J_{(k\beta)}^{(-)}(v,\cdot) & = & 
-i \lim_{\staccrel{\scriptstyle \epsilon\rightarrow 0}{\scriptstyle \epsilon>0}}
\left\{\cG^{(-)}(k\beta+\epsilon+iv,\cdot)- \cG^{(-)}(k\beta-\epsilon+iv,\cdot) \right\},~~~~
(v\leq 0,\; k\in\Z). \label{diciassetteb}
\end{eqnarray}
Let us note that these definitions are well--posed and appropriate because, as we 
shall see in the following theorem, $\cG^{(+)}(\tau,\cdot)$ and $\cG^{(-)}(\tau,\cdot)$ 
are holomorphic in the cut--domains $\cI_- \cup [\cI_+ \setminus \Cir{\Xi}_+]$ and 
$\cI_+ \cup [\cI_- \setminus \Cir{\Xi}_-]$, respectively.
Moreover, we suppose hereafter that the slow--growth condition (\ref{dodici}) extends 
to the discontinuities $J_{(k\beta)}^{(\pm)}(v,\cdot)$, that turn out to be 
``tempered functions''~\cite{Bremermann}.
Finally, in view of the periodicity properties of $\cG(\tau,\cdot)$, it
is sufficient to consider only the strip, in the $\tau$--plane, defined
by $-a\leq u \leq 2\beta-a$ $(0<a<\beta)$, $v\in\R$ (see Fig.~\ref{figura_1}).

We then introduce the Laplace transforms of the jump functions across the cuts 
located at $\Real\tau = 0$, and at $\Real\tau = \beta$; i.e.
\begin{eqnarray}
\tJ_{(0)}^{(+)} (\zeta,\cdot) & = &
\int_0^{+\infty} J_{(0)}^{(+)} (v, \cdot) \, e^{-\zeta v} \, dv,
~~~~~(\zeta=\xi+i\eta,\, \Real\zeta>0), \label{diciottoa} \\
\tJ_{(0)}^{(-)} (\zeta,\cdot) & = &
\int_{-\infty}^0 J_{(0)}^{(-)} (v, \cdot) \, e^{-\zeta v} \, dv,
~~~~~(\Real\zeta<0), \label{diciottob} \\
\tJ_{(\beta)}^{(+)} (\zeta,\cdot) & = &
\int_0^{+\infty} J_{(\beta)}^{(+)}(v, \cdot)\,e^{-\zeta v}\,dv,
~~~~~(\Real\zeta>0), \label{diciottoc} \\
\tJ_{(\beta)}^{(-)} (\zeta,\cdot) & = &
\int_{-\infty}^0 J_{(\beta)}^{(-)} (v, \cdot) \, e^{-\zeta v}\,dv, 
~~~~~(\Real\zeta<0). \label{diciottod}
\end{eqnarray}
We can state the following theorem.

\begin{theorem}
\label{theorem1}
If the functions $\cG(\tau,\cdot)$ and $J_{(k\beta)}^{(\pm)}(v,\cdot)$ 
satisfy the slow--growth condition {\rm(}\ref{dodici}{\rm)} uniformly in 
$\Cir{\Pi}_\tau = \C\setminus (\Cir{\Xi}_+ \cup \Cir{\Xi}_-)$ up to the closure, 
the following properties hold true:
\item[i{\rm)}]
The function $\cG^{(+)} (\tau,\cdot)$ {\rm (}respectively $\cG^{(-)}(\tau,\cdot)${\rm)}
is holomorphic in the cut--domain $\cI_- \cup [\cI_+ \setminus \Cir{\Xi}_+]$
{\rm(}respectively $\cI_+ \cup [\cI_- \setminus \Cir{\Xi}_-]${\rm)}.
\item[ii-a{\rm)}]
The Laplace transforms $\tJ_{(0)}^{(+)} (\zeta,\cdot)$ and
$\tJ_{(\beta)}^{(+)} (\zeta,\cdot)$ are holomorphic in the half--plane $\Real\zeta > 0$.
The Laplace transforms $\tJ_{(0)}^{(-)} (\zeta,\cdot)$ and $\tJ_{(\beta)}^{(-)} (\zeta,\cdot)$
are holomorphic in the half--plane $\Real\zeta < 0$.
\item[ii-b{\rm)}]
$\tJ_{(0)}^{(+)} (\zeta,\cdot)$ and $\tJ_{(\beta)}^{(+)} (\zeta,\cdot)$ belong to
the Hardy space $\HH^2\left(\C_{(\delta)}^{(+)}\right)$, where
$\C_{(\delta)}^{(+)}=\{\zeta\in\C\,|\,\Real\zeta > \delta ,\, \delta\geq\epsilon>0\}$.
$\tJ_{(0)}^{(-)} (\zeta,\cdot)$ and $\tJ_{(\beta)}^{(-)} (\zeta,\cdot)$ belong to the
Hardy space $\HH^2\left (\C_{(\delta)}^{(-)}\right )$, where
$\C_{(\delta)}^{(-)}=\{\zeta\in\C\,|\,\Real\zeta<\delta ,\, \delta\geq\epsilon>0\}$.
\item[iii-a{\rm)}]
In the case of the boson statistics the symmetric combinations
$\tcG^{(+,b)}(\zeta,\cdot) :\equiv \tJ_{(0)}^{(+)}(\zeta,\cdot)+\tJ_{(\beta)}^{(+)}(\zeta,\cdot)$ and
$\tcG^{(-,b)}(\zeta,\cdot) :\equiv \tJ_{(0)}^{(-)}(\zeta,\cdot)+\tJ_{(\beta)}^{(-)}(\zeta,\cdot)$ 
interpolate uniquely the Fourier coefficients $\cG^{(+)}_{2m}(\cdot)$ and $\cG^{(-)}_{2m}(\cdot)$
respectively {\rm (}hereafter the superscript $(b)$ stands for the boson statistics{\rm )}. 
Let $\zeta_m = 2m\pi/\beta$, then the following Froissart--Gribov--type equalities hold:
\begin{eqnarray}
\tcG^{(+,b)}(\zeta_m,\cdot) = 
\tJ_{(0)}^{(+)} (\zeta_m,\cdot)+\tJ_{(\beta)}^{(+)} (\zeta_m,\cdot) & = &
\cG^{(+)}_{2m}(\cdot),~~~~~~(m=1,2,3,\ldots), \label{diciannovea} \\
\tcG^{(-,b)}(\zeta_m,\cdot) = 
\tJ_{(0)}^{(-)} (\zeta_m,\cdot)+\tJ_{(\beta)}^{(-)} (\zeta_m,\cdot) & = &
\cG^{(-)}_{2m}(\cdot),~~~~~~(m=-1,-2,-3,\ldots). \label{diciannoveb}
\end{eqnarray}
\item[iii-b{\rm)}]
In the case of the fermion statistics the antisymmetric combinations
$\tcG^{(+,f)}(\zeta,\cdot) :\equiv \tJ_{(0)}^{(+)} (\zeta,\cdot) - \tJ_{(\beta)}^{(+)} (\zeta,\cdot)$ and
$\tcG^{(-,f)}(\zeta,\cdot) :\equiv \tJ_{(0)}^{(-)} (\zeta,\cdot) - \tJ_{(\beta)}^{(-)} (\zeta,\cdot)$ 
interpolate uniquely the Fourier coefficients $\cG^{(+)}_{2m+1}(\cdot)$ and $\cG^{(-)}_{2m+1}(\cdot)$
respectively {\rm (}hereafter the superscript $(f)$ stands for the fermion statistics{\rm )}.
Let $\zeta_m = (2m+1)\pi/\beta$, then the following Froissart--Gribov--type equalities hold:
\begin{eqnarray}
\tcG^{(+,f)}(\zeta_m,\cdot) = 
\tJ_{(0)}^{(+)} (\zeta_m,\cdot)-\tJ_{(\beta)}^{(+)} (\zeta_m,\cdot) & = &
\cG^{(+)}_{2m+1}(\cdot),~~~~~~(m=0,1,2,3,\ldots), \label{ventia} \\
\tcG^{(-,f)}(\zeta_m,\cdot) = 
\tJ_{(0)}^{(-)} (\zeta_m,\cdot)-\tJ_{(\beta)}^{(-)} (\zeta_m,\cdot) & = &
\cG^{(-)}_{2m+1}(\cdot),~~~~~~(m=-1,-2,-3,\ldots). \label{ventib}
\end{eqnarray}
\end{theorem}

\begin{proof}
(i) In view of the Riemann--Lebesgue theorem, and since 
$\cG^{(+)}(\tau,\cdot)\in L_1[-\beta,\beta]$, the Fourier coefficients 
$\cG^{(+)}_n(\cdot)$ tend to zero as $n\rightarrow\infty$. From expansion 
(\ref{quindicia}) we have for all $\tau=u+iv$, with $v<0$:
\beq
\label{ventisei}
|\cG^{(+)} (\tau,\cdot)|=\left|\frac{1}{2\beta}\sum_{n=0}^{+\infty}\cG^{(+)}_n(\cdot)
e^{-i\zeta_n\tau}\right| \leq K \sum_{n\geq 0} e^{\zeta_n v},
\eeq
where $K=\int_{-\beta}^{\beta}|\cG(\tau,\cdot)|\,d\tau$.
The series $\sum_{n\geq 0}^{+\infty}e^{\zeta_n v}$ converges uniformly in any 
domain compactly contained in the half--plane $\Imag\tau<0$. In view of the 
Weierstrass theorem on the uniformly convergent series of analytic functions, 
we can conclude that $\cG^{(+)}(\tau,\cdot)$ is holomorphic in the half--plane 
$\Imag\tau<0$. By using analogous arguments we can prove that $\cG^{(-)}(\tau,\cdot)$ 
is holomorphic in the half--plane $\Imag\tau>0$.  Furthermore, we know from 
Hypothesis a) that $\cG(\tau,\cdot)=\cG^{(+)}(\tau,\cdot)+\cG^{(-)}(\tau,\cdot)$ 
is holomorphic in the strips $k\beta<u<(k+1)\beta$ ($k\in\Z,\,v\in\R$), and 
continuous at the boundaries of the strips. We can conclude that 
$\cG^{(+)}(\tau,\cdot)$ is holomorphic in the cut--domain
$\cI_- \cup [\cI_+ \setminus \Cir{\Xi}_+]$, and $\cG^{(-)}(\tau,\cdot)$ 
is holomorphic in the cut--domain
$\cI_+ \cup [\cI_- \setminus \Cir{\Xi}_-]$. \\
(ii) Property (ii, a) follows easily from the assumption of ``temperateness'' of 
the jump functions~\cite{Bremermann}. For what concerns property (ii, b) we limit 
ourselves to prove that $\tJ_{(0)}^{(+)} (\zeta,\cdot)$ belongs to the Hardy 
space $\HH^2\left (\C_{(\delta)}^{(+)}\right )$, since the remaining part of the 
statement can be proved analogously. To this purpose, we rewrite the Laplace 
transform (\ref{diciottoa}) in the following form:
\beq
\label{ventuno}
\int_0^{+\infty} \left (J_{(0)}^{(+)} (v,\cdot) e^{-\delta v}\right ) e^{-\zeta' v}\, dv :\equiv
\tJ_{(0)(\delta)}^{(+)} (\zeta',\cdot),~~~~~~(\Real\zeta' > 0),
\eeq
where $\Real\zeta' = \Real\zeta - \delta$ $(\delta\geq\epsilon>0)$. In view of the 
slow--growth property of $J_{(0)}^{(+)} (v,\cdot)$, we can then say that the function 
$J_{(0)}^{(+)} (v,\cdot) \exp(-\delta v)$ belongs to the intersection 
$L^1[0,+\infty) \cap L^2[0,+\infty)$.
Then, thanks to the Paley--Wiener theorem, we can conclude (returning to the variable 
$\zeta$) that $\tJ_{(0)}^{(+)} (\zeta,\cdot)$ belongs to the Hardy space 
$\HH^2\left (\C_{(\delta)}^{(+)}\right )$ (see Ref.~\cite{Hoffman}).
Accordingly, $\tJ_{(0)}^{(+)} (\zeta,\cdot)$ tends uniformly to zero as $\zeta$ 
tends to infinity inside any fixed half--plane $\Real\zeta\geq\delta'>\delta$. 
In particular, $\tJ_{(0)}^{(+)} (\zeta_n,\cdot)$, with $\zeta_n=n\pi/\beta$ $(n=1,2,\ldots)$, 
tends to zero as $n\rightarrow\infty$. \\
(iii) We introduce the integral $I^{(+)}_\gamma$ defined as follows (this method 
has been introduced by Bros and Buchholz~\cite{Bros1}, and will be developed in a more detailed 
form in~\cite{Bros3} within the general framework of QFT):
\beq
\label{ventidue}
I^{(+)}_\gamma (\zeta,\cdot)=\int_\gamma \cG^{(+)}(\tau,\cdot)\,e^{i\zeta\tau}\,d\tau,
\eeq
where the path $\gamma$ encloses both the cuts located at $u=0$, $v\geq 0$ and at 
$u=\beta$, $v\geq 0$ (see Fig.~\ref{figura_1}). In view of the slow--growth condition 
(\ref{dodici}), this integral is well--defined.
By choosing as integration path a pair of contours $(\gamma_0,\gamma_\beta)$ 
enclosing respectively the cuts at $u=0$, $v\geq 0$ and at $u=\beta$, $v\geq 0$, 
and then flattening them (in a folded way) onto the cuts (see Fig.~\ref{figura_1}), 
we obtain:
\beq
\label{ventitre}
I^{(+)}_{(\gamma_0\cup\gamma_\beta)}(\zeta,\cdot)=
\int_0^{+\infty} J_{(0)}^{(+)} (v,\cdot) \,e^{-\zeta v}\, dv +
e^{i\zeta\beta}\int_0^{+\infty} J_{(\beta)}^{(+)} (v,\cdot) \,e^{-\zeta v}\, dv =
\tJ_{(0)}^{(+)} (\zeta,\cdot) + e^{i\zeta\beta} \tJ_{(\beta)}^{(+)} (\zeta,\cdot).
\eeq
Next, we choose the path $\gamma_{a_\epsilon}$, whose support is:
$]-a+i\infty,-a]\cup[-a,-\epsilon]\cup[\gamma_\epsilon^{(0)}]\cup[\epsilon,\beta-\epsilon]\cup
[\gamma_\epsilon^{(\beta)}]\cup[\beta+\epsilon,2\beta-a]\cup[2\beta-a,2\beta-a+i\infty[$,
where $\gamma_\epsilon^{(0)}$ and $\gamma_\epsilon^{(\beta)}$ are half--circles 
turning around the points $\tau=0$ and $\tau=\beta$, respectively (see Fig.~\ref{figura_1}).
By taking into account the $2\beta$--periodicity of $\cG^{(+)} (\tau,\cdot)$, we get, for
$\zeta=\zeta_n=n\pi/\beta$, ($n=1,2,\ldots$):
\beq
\label{ventiquattrotris}
\lim_{\epsilon\rightarrow 0} I^{(+)}_{\gamma_{a_\epsilon}}(\zeta_n,\cdot)=
\int_{-a}^{2\beta-a} \cG^{(+)} (\tau,\cdot) \, e^{i\zeta_n\tau}\,d\tau = \cG^{(+)}_n(\cdot).
\eeq
Then, from the Cauchy distortion argument, we have
$I^{(+)}_{\gamma_0 \cup \gamma_\beta}(\zeta_n,\cdot)=
\lim_{\epsilon\rightarrow 0} I^{(+)}_{\gamma_{a_\epsilon}}(\zeta_n,\cdot)$, that is
\beq
\label{venticinque}
\tJ_{(0)}^{(+)} (\zeta_n,\cdot)+e^{i\zeta_n\beta}\,\tJ_{(\beta)}^{(+)} (\zeta_n,\cdot) =
\cG^{(+)}_n(\cdot).
\eeq
We now distinguish two cases:
\begin{itemize}
\item[1)] $n$ even: $n=2m$, $\zeta_m=2m\pi/\beta$ $(m=1,2,\ldots)$; then from 
(\ref{venticinque}) we obtain equalities (\ref{diciannovea}).
\item[2)] $n$ odd: $n=2m+1$, $\zeta_m=(2m+1)\pi/\beta$ $(m=0,1,2,\ldots)$;
then from (\ref{venticinque}) we obtain equalities (\ref{ventia}).
\end{itemize}
We have thus obtained two combinations (symmetric and antisymmetric, respectively) 
that interpolate the Fourier coefficients $\cG^{(+)}_n(\cdot)$. The uniqueness of 
the interpolation is guaranteed by the Carlson theorem~\cite{Boas} that can be 
applied since $\tJ_{(0)}^{(+)} (\zeta,\cdot)$ and $\tJ_{(\beta)}^{(+)} (\zeta,\cdot)$ 
belong to the Hardy space $\HH^2\left (\C_{(\delta)}^{(+)}\right )$.
Proceeding with analogous arguments applied to $\cG^{(-)} (\tau,\cdot)$ equalities
(\ref{diciannoveb}) and (\ref{ventib}) are obtained.
\qed
\end{proof}

In conclusion, we can say that the thermal Green functions present a double 
analytic structure involving the analyticity properties in the $\tau=u+iv$ and 
$\zeta=\xi+i\eta$ planes. The $2\beta$--periodic function $\cG^{(+)}(\tau,\cdot)$ 
(resp. $\cG^{(-)}(\tau,\cdot)$) is analytic in the cut--domain 
$\cI_- \cup [\cI_+ \setminus \Cir{\Xi}_+]$
(resp. $\cI_+ \cup [\cI_- \setminus \Cir{\Xi}_-]$);
its Fourier coefficients can be uniquely interpolated (in the sense of the Carlson 
theorem), and are the restriction to the appropriate Matsubara energies of a function
$\tcG^{(+,b-f)}(\zeta,\cdot)$ (resp. $\tcG^{(-,b-f)}(\zeta,\cdot)$), analytic in 
the half--plane $\Real\zeta>0$ (resp. $\Real\zeta<0$).
It is straightforward to verify that the jump function $J^{(+)}_{(0)}(v,\cdot)$ 
coincides with the retarded Green function, and $J^{(-)}_{(0)}(v,\cdot)$ coincides 
with the advanced one; analogously, putting $i\,\zeta=\omega$, we can identify 
$\tcG^{(+,b-f)}(\zeta,\cdot)$ and $\tcG^{(-,b-f)}(\zeta,\cdot)$ respectively with 
the retarded and advanced Green functions in the energy variable $\omega$ conjugate 
to the real time $t$.

\section{Representation of the Jump Function in Terms of an Infinite Set of Fourier Coefficients}
\label{se:representation}
First let us consider a system of bosons; since $n$ is even, i.e. $n=2m$, 
$\zeta_m=(2m\pi)/\beta$, ($m=0,1,2,\ldots$), we have:
\beq
\label{Idue}
\tcG^{(+,b)}\left(\frac{2m\pi}{\beta},\cdot\right)=
2\int_0^\beta \cG^{(+)}(\tau,\cdot)\,e^{i\frac{2m\pi}{\beta}\tau}\,d\tau\,.
\eeq
Next, recalling that $\cG^{(+)}(\tau,\cdot)$ is $\beta$--periodic, we can write 
also the following Fourier expansion:
\begin{eqnarray}
\cG^{(+)}(\tau,\cdot) & = & 
\frac{1}{\beta}\sum_{m=0}^\infty \tcG^{(+,b)}_{(\beta)}\left(\frac{2m\pi}{\beta},\cdot\right)
e^{-i\frac{2m\pi}{\beta}\tau}, \label{Itrea} \\
\tcG^{(+,b)}_{(\beta)}\left(\frac{2m\pi}{\beta},\cdot\right) &=& 
\int_0^\beta \cG^{(+)}(\tau,\cdot)\, e^{i\frac{2m\pi}{\beta}\tau}\, d\tau = 
\frac{1}{2}\, \tcG^{(+,b)}\left(\frac{2m\pi}{\beta},\cdot\right). \label{Itreb}
\end{eqnarray}
Finally, putting $\beta=2\pi$, formulae (\ref{Itrea}), (\ref{Itreb}) can be rewritten
in the more convenient form:
\begin{eqnarray}
\cG^{(+)}(\tau,\cdot) & = & \frac{1}{2\pi}\sum_{m=0}^\infty \tcG^{(+,b)}_{(2\pi)}(m,\cdot)
e^{-im\tau}, \label{Iquattroa} \\
\tcG^{(+,b)}_{(2\pi)}(m,\cdot) &=& \int_0^{2\pi} \cG^{(+)}(\tau,\cdot)\,
e^{im\tau}\, d\tau = \frac{1}{2} \,\tcG^{(+,b)}(m,\cdot). \label{Iquattrob}
\end{eqnarray}
Recalling once again the $\beta$--periodicity of the function $\cG^{(+)}(\tau,\cdot)$, 
we write now the Froissart--Gribov equalities (\ref{diciannovea}) as
\beq
\label{Icinque}
\tcG^{(+,b)}(m,\cdot) = \tJ^{(+,b)}_{(0)} (m,\cdot) + \tJ^{(+,b)}_{(2\pi)} (m,\cdot) =
2 \tJ^{(+,b)}_{(0)} (m,\cdot) = 2 \tcG^{(+,b)}_{(2\pi)} (m,\cdot),~~~~(m=1,2,3,\ldots).
\eeq
It is now convenient to introduce an auxiliary function $J^{(b)}_* (v,\cdot)$, defined as follows:
\beq
\label{Isei}
J^{(b)}_* (v,\cdot) = e^{-v}\, J^{(+,b)}_{(0)} (v,\cdot),~~~~~~ (v\in\R^+),
\eeq
and the corresponding Laplace transform:
\beq
\label{Isette}
\tJ^{(b)}_* (\zeta,\cdot) = \int_0^{+\infty} J^{(b)}_* (v,\cdot) \, e^{-\zeta v}\,dv,
~~~~~~(\zeta=\xi+i\eta, \Real\zeta > -1+\delta,\,\delta\geq\epsilon>0).
\eeq
It is straightforward to prove, via the Paley--Wiener theorem, that
$\tJ^{(b)}_* (\zeta,\cdot)$ belongs to the Hardy space 
$\HH^2\left (\C_{(-1+\delta)}^{(+)}\right )$, where 
$\C_{(-1+\delta)}^{(+)}=\{\zeta\in\C\,|\,\Real\zeta>-1+\delta,\,\delta\geq\epsilon>0\}$.
Next, the Froissart--Gribov equalities (\ref{Icinque}) can be rewritten as
\beq
\label{Iotto}
\tJ^{(b)}_* (m,\cdot) = \tcG^{(+,b)}_{(2\pi)}(m+1,\cdot)~,~~~~~(m=0,1,2,\ldots).
\eeq
Then we can prove the following lemma.

\begin{lemma}
\label{lemma1}
The function $\tJ^{(b)}_* (-1/2+i\eta,\cdot)$, $(\eta\in\R)$ can be represented by
the following series, that converges in the sense of the $L^2$--norm:
\beq
\label{Inove}
\tJ^{(b)}_* \left(-\frac{1}{2}+i\eta,\cdot\right)=\sum_{\ell=0}^\infty c_\ell\psi_\ell(\eta),
\eeq
$\psi_\ell(\eta)$ denoting the Pollaczek functions defined by
\beq
\label{Idieci}
\psi_\ell(\eta) = \frac{1}{\sqrt\pi}\,\Gamma\left(\frac{1}{2}+i\eta\right)P_\ell(\eta),
\eeq
$\Gamma$ being the Euler gamma function, and $P_\ell$ the Pollaczek polynomials~\cite{Bateman,Szego}.
The coefficients $c_\ell$ are given by:
\beq
\label{Iundici}
c_\ell = 2\sqrt\pi\sum_{m=0}^\infty\frac{(-1)^m}{m!}\tcG^{(+,b)}_{(2\pi)}(m+1,\cdot)
P_\ell\left[-i\left(m+\frac{1}{2}\right)\right].
\eeq
\end{lemma}

\begin{figure}[t]
\centerline{\psfig{file=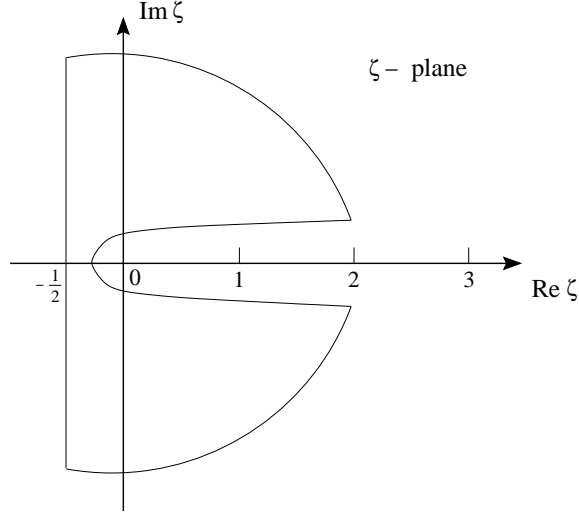}}
\caption{Integration path for the evaluation of integral (\ref{Isedici})}
\label{figura_2}
\end{figure}

\begin{proof}
The Pollaczek polynomials $P_\ell^{(\alpha)}(\eta)$, ($\eta\in\R$), are 
orthogonal in $L^2(-\infty,+\infty)$ with weight function (see Refs.~\cite{Bateman,Szego}):
\beq
\label{Idodici}
w(\eta) = \frac{1}{\pi}\,2^{(2\alpha-1)} \left|\Gamma(\alpha+i\eta)\right|^2.
\eeq
For $\alpha=1/2$, the orthogonality property reads:
\beq
\label{Itredici}
\int_{-\infty}^{+\infty}w(\eta)P_\ell^{(1/2)}(\eta)P_{\ell'}^{(1/2)}(\eta)\,d\eta=
\delta_{\ell,\ell'}~,~~~
\left( w(\eta)=\frac{1}{\pi}\left|\Gamma\left(\frac{1}{2}+i\eta\right)\right|^2\right),
\eeq
(in the following, when $\alpha=1/2$, we omit the index $\alpha$ in the notation). Next, 
we introduce the following functions, that will be called Pollaczek functions 
(of index $\alpha=1/2$):
\beq
\label{Iquattordici}
\psi_\ell(\eta) = \frac{1}{\sqrt\pi}\,\Gamma\left(\frac{1}{2}+i\eta\right)P_\ell(\eta),
\eeq
which form a complete basis in $L^2(-\infty,+\infty)$~\cite{Itzykson}.
Since $\tJ^{(b)}_* (\zeta,\cdot)$ belongs to the Hardy space 
$\HH^2\left(\C_{(-1+\delta)}^{(+)}\right)$, then $\tJ^{(b)}_*(-1/2+i\eta,\cdot)$ 
($\eta\in\R$) belongs to $L^2(-\infty,+\infty)$. Therefore, we may expand
$\tJ^{(b)}_*(-1/2+i\eta,\cdot)$ in terms of Pollaczek functions as follows:
\beq
\label{Iquindici}
\tJ^{(b)}_*\left(-\frac{1}{2}+i\eta,\cdot\right) = \sum_{\ell=0}^\infty c_\ell\psi_\ell(\eta),
\eeq
where the series at the r.h.s. of (\ref{Iquindici}) converges to
$\tJ^{(b)}_*(-1/2+i\eta,\cdot)$ in the sense of the $L^2$--norm. From (\ref{Iquindici}) we get
\beq
\label{Isedici}
c_\ell = \frac{1}{\sqrt\pi}\int_{-\infty}^{+\infty}\tJ^{(b)}_*\left(-\frac{1}{2}+i\eta,\cdot\right)
\Gamma\left(\frac{1}{2}-i\eta\right)P_\ell(\eta)\,d\eta.
\eeq
The integral at the r.h.s. of (\ref{Isedici}) can be evaluated by the contour 
integration method along the path shown in Fig.~\ref{figura_2}, and taking into 
account the asymptotic behaviour of the gamma function given by the Stirling
formula. We obtain:
\beq
\label{Idiciassette}
c_\ell = 2\sqrt\pi\sum_{m=0}^\infty\frac{(-1)^m}{m!}\,\tJ^{(b)}_* (m,\cdot)
P_\ell\left[-i\left(m+\frac{1}{2}\right)\right].
\eeq
Finally, from (\ref{Iotto}), (\ref{Iquindici}) and (\ref{Idiciassette}) the proof 
of the lemma follows.
\qed
\end{proof}

From (\ref{Isette}), when $\zeta=-1/2+i\eta$ ($\eta\in\R$), we have:
\beq
\label{Idiciotto}
\tJ^{(b)}_*\left(-\frac{1}{2}+i\eta,\cdot\right) = 
\int_0^{+\infty}J^{(b)}_*(v,\cdot)\,e^{v/2} e^{-i\eta v}\,dv.
\eeq
The r.h.s. of (\ref{Idiciotto}) is the Fourier transform of $J^{(b)}_*(v,\cdot) e^{v/2}$.
Noting that $\tJ^{(b)}_*(-1/2+i\eta,\cdot)$ belongs to $L^2(-\infty,+\infty)$, but 
not necessarily to $L^1(-\infty,+\infty)$, the inversion of the Fourier transform 
(\ref{Idiciotto}) holds only as a limit in the mean order two, and can be written as follows:
\beq
\label{Idiciannove}
J^{(b)}_*(v,\cdot)\, e^{v/2} = \staccrel{{\rm l.i.m.}}{\eta_0\rightarrow +\infty}
\left (\frac{1}{2\pi}\int_{-\eta_0}^{\eta_0}
\tJ^{(b)}_*\left(-\frac{1}{2}+i\eta,\cdot\right) e^{i\eta v}\,d\eta\right ),~~~~~(v\in\R^+).
\eeq
Then, we can prove the following lemma.

\begin{lemma}
\label{lemma2}
The function $J^{(b)}_*(v,\cdot) e^{v/2}$ can be represented by the following expansion
that converges in the sense of the $L^2$--norm:
\beq
\label{Iventi}
e^{v/2}J^{(b)}_*(v,\cdot) = \sum_{\ell=0}^\infty a_\ell \Phi_\ell (v),~~~~~(v\in\R^+),
\eeq
where the coefficients $a_\ell$ are given by:
\beq
\label{Iventuno}
a_\ell = \sqrt 2\sum_{m=0}^\infty\frac{(-1)^m}{m!}\,\tcG^{(+,b)}_{(2\pi)}(m+1,\cdot)\,
P_\ell\left[-i\left(m+\frac{1}{2}\right)\right],
\eeq
$P_\ell$ being the Pollaczek polynomials, and the functions $\Phi_\ell(v)$ are given by
\beq
\label{Iventidue}
\Phi_\ell(v)=i^\ell\sqrt 2 \, L_\ell(2e^{-v})\, e^{-e^{-v}} e^{-v/2},
\eeq
$L_\ell$ being the Laguerre polynomials.
\end{lemma}

\begin{proof}
Let us observe that
\beq
\label{Iventitre}
\Gamma\left(\frac{1}{2}+i\eta\right)=\int_0^{+\infty}e^{-t}\,t^{(i\eta-1/2)}\,dt =
\int_{-\infty}^{+\infty}e^{-e^{-v}} e^{-v/2}\, e^{-i\eta v}\,dv=
{\cal F}\left\{e^{-e^{-v}} e^{-v/2}\right\},
\eeq
where ${\cal F}$ denotes the Fourier integral operator. Let us note that the function
$\exp(-e^{-v}) e^{-v/2}$ belongs to $S^{\infty}(\R)$, i.e. the Schwartz space of the 
$C^\infty(\R)$ functions that, together with all their derivatives, tend to zero, 
for $|v|$ tending to $+\infty$, faster than any negative power of $|v|$.
Therefore, we can write (see formula (\ref{Iquattordici})):
\beq
\label{Iventiquattro}
\psi_\ell(\eta) = 
\frac{1}{\sqrt\pi}\,{\cal F}\left\{P_\ell\left(-i\frac{d}{dv}\right)\left[e^{-e^{-v}}e^{-v/2}\right]
\right\}.
\eeq
Substituting in expansion (\ref{Inove}) to the Pollaczek functions their representation
(\ref{Iventiquattro}), we obtain:
\beq
\label{Iventicinque}
\tJ^{(b)}_*(-\frac{1}{2}+i\eta,\cdot)=\sum_{\ell=0}^\infty c_\ell \left\{
\frac{1}{\sqrt\pi}\,{\cal F}\left[P_\ell\left(-i\frac{d}{dv}\right)\left(e^{-e^{-v}} 
e^{-v/2}\right)\right]\right\}.
\eeq
Let us now apply the operator ${\cal F}^{-1}$ to the r.h.s. of (\ref{Iventicinque}). 
If we exchange the integral operator ${\cal F}^{-1}$ with the sum, and this is 
legitimate within the $L^2$--norm convergence, we obtain:
\beq
\label{Iventisei}
{\cal F}^{-1}\sum_{\ell=0}^\infty c_\ell \left\{
\frac{1}{\sqrt\pi}\,{\cal F}\left[P_\ell\left(-i\frac{d}{dv}\right)
\left(e^{-e^{-v}} e^{-v/2}\right)\right]\right\} =
\sum_{\ell=0}^\infty c_\ell \left\{\frac{1}{\sqrt\pi}\left[P_\ell\left(-i\frac{d}{dv}\right)
\left(e^{-e^{-v}} e^{-v/2}\right)\right]\right\}.
\eeq
Finally, recalling formula (\ref{Idiciannove}), we obtain the following expansion 
for the function $J^{(b)}_*(v,\cdot)e^{v/2}$:
\beq
\label{Iventisette}
e^{v/2}J^{(b)}_*(v,\cdot) = 
\sum_{\ell=0}^\infty\frac{c_\ell}{\sqrt\pi}\,P_\ell\left(-i\frac{d}{dv}\right)
\left(e^{-e^{-v}} e^{-v/2}\right),
\eeq
whose convergence is in the sense of the $L^2$--norm. It can be easily 
verified that~\cite{DeMicheli}
\beq
\label{Iventotto}
\sqrt 2 P_\ell\left(-i\frac{d}{dv}\right)\left(e^{-e^{-v}} e^{-v/2}\right)=
i^\ell\sqrt 2 \, L_\ell\left(2e^{-v}\right) e^{-e^{-v}}e^{-v/2},
\eeq
where $L_\ell$ denotes the Laguerre polynomials.

It can be checked that the polynomials ${\cal L}_\ell(v)=i^\ell\sqrt 2 \, L_\ell(2e^{-v})$ 
are a set of polynomials orthonormal on the real line with weight function 
$w(v)=\exp(-v)\exp(-2e^{-v})$, and, consequently, the set of functions $\Phi_\ell(v)$, 
defined by formula (\ref{Iventidue}), forms an orthonormal basis in $L^2(-\infty,+\infty)$.

Finally, from (\ref{Iventisette}) we obtain:
\beq
\label{Iventinove}
e^{v/2}J^{(b)}_*(v,\cdot) =
\sum_{\ell=0}^\infty a_\ell\left\{i^\ell\sqrt 2 \, L_\ell(2e^{-v})\, e^{-e^{-v}} e^{-v/2}\right\} =
\sum_{\ell=0}^\infty a_\ell \Phi_\ell(v),~~~~(v\in\R^+),
\eeq
where $a_\ell=c_\ell/\sqrt{2\pi}$, and the functions $\Phi_\ell(v)$ are given 
by formula (\ref{Iventidue}).
\qed
\end{proof}

We now introduce the weighted $L^2$--space $L^2_{(w)}[0,+\infty)$, whose norm is defined by:
\beq
\label{new_1}
\left\| f \right\|_{L^2_{(w)}[0,+\infty)} =
\left( \int_0^{+\infty} w(v) \left | f(v)\right |^2\,dv \right )^{1/2},
\eeq
$w(v)$ being a weight function which will be specified in the following. Then we can prove
the following result.

\begin{theorem}
\label{theorem2}
The jump function $J^{(+,b)}_{(0)}(v,\cdot)$ can be represented by the following expansion:
\beq
\label{Itrenta}
J^{(+,b)}_{(0)}(v,\cdot)= e^{v/2}\sum_{\ell=0}^\infty a_\ell\Phi_\ell(v),~~~~~(v \in\R^+),
\eeq
which converges in the sense of the $L^2_{(w)}[0,+\infty)$--norm, with weight function 
$w(v)=e^{-v},\,(v\in\R^+)$.
\end{theorem}

\begin{proof}
We can write:
\begin{eqnarray}
\label{new_2}
\left\|J^{(+,b)}_{(0)}(v,\cdot)-e^{v/2}\sum_{\ell=0}^L 
a_\ell\Phi_\ell(v)\right\|_{L^2_{(w)}[0,+\infty)} &=&
\left(\int_0^{+\infty}e^{-v}\left|J^{(+,b)}_{(0)}(v,\cdot)-e^{v/2}\sum_{\ell=0}^L 
a_\ell\Phi_\ell(v)\right|^2\,dv\right)^{1/2} \nonumber \\
&=&\left(\int_0^{+\infty}\left|e^{v/2}J^{(b)}_{*}(v,\cdot)-\sum_{\ell=0}^L
a_\ell\Phi_\ell(v)\right|^2\,dv\right)^{1/2}.
\end{eqnarray}
In view of Lemma~\ref{lemma2} we can thus state that:
\beq
\label{new_3}
\lim_{L\rightarrow\infty}\left\|J^{(+,b)}_{(0)}(v,\cdot)-e^{v/2}\sum_{\ell=0}^L 
a_\ell\Phi_\ell(v)\right\|_{L^2_{(w)}[0,+\infty)}=0,
\eeq
that proves the statement.
\qed
\end{proof}

Consider now a system of fermions. In this case the function $\cG^{(+)}(\tau,\cdot)$
is antiperiodic with period $\beta$. 
Then, if we put $\zeta_m=(2m+1)\pi/\beta$ ($m=0,1,2,\ldots$) and $\beta=2\pi$, we have 
the following expansion:
\begin{eqnarray}
\cG^{(+)}(\tau,\cdot) &=& \frac{1}{2\pi}\sum_{m=0}^\infty\tcG^{(+,f)}_{(2\pi)}
\left(m+\frac{1}{2},\cdot\right) e^{-i(m+1/2)\tau}, \label{Itrentacinquea} \\
\tcG^{(+,f)}_{(2\pi)}\left(m+\frac{1}{2},\cdot\right)&=& \int_0^{2\pi}\cG^{(+)}(\tau,\cdot)\,
e^{i(m+1/2)\tau}\,d\tau = 
\frac{1}{2}\,\tcG^{(+,f)}\left(m+\frac{1}{2},\cdot\right). \label{Itrentacinqueb}
\end{eqnarray}
Recalling once again the antiperiodicity of $\cG^{(+)}(\tau,\cdot)$, we write the 
Froissart--Gribov equalities (\ref{ventia}) in the following form:
\begin{eqnarray}
\label{Itrentasei}
\tcG^{(+,f)}\left(m+\frac{1}{2},\cdot\right) &=&
\tJ^{(+,f)}_{(0)}\left(m+\frac{1}{2},\cdot\right)-
\tJ^{(+,f)}_{(2\pi)}\left(m+\frac{1}{2},\cdot\right) \nonumber \\
&=& 2\,\tJ^{(+,f)}_{(0)}\left(m+\frac{1}{2},\cdot\right)=
2\,\tcG^{(+,f)}_{(2\pi)}\left(m+\frac{1}{2},\cdot\right), ~~~~~(m=0,1,2,\ldots).
\end{eqnarray}
We can now proceed in a way strictly analogous to that followed in the case of bosons. We put:
$J^{(f)}_*(v,\cdot)=e^{-v}J^{(+,f)}_{(0)}(v,\cdot)$ and, accordingly,
$\tJ^{(f)}_*(\zeta,\cdot)=\int_0^{+\infty}J^{(f)}_*(v,\cdot)e^{-\zeta v}dv$ 
($\zeta=\xi+i\eta, \Real\zeta\equiv\xi > -1+\delta$, $\delta\geq\epsilon>0$).
Then, the Froissart--Gribov equalities (\ref{Itrentasei}) now read:
\beq
\label{Itrentasette}
\tJ^{(f)}_*\left(m+\frac{1}{2},\cdot\right)=\tcG^{(+,f)}_{(2\pi)}\left(m+\frac{3}{2},\cdot\right),
~~~~~(m=0,1,2,\ldots).
\eeq
We can now state the following theorem.

\begin{theorem}
\label{theorem3}
i{\rm)} The function $\tJ^{(f)}_*(i\eta,\cdot)$, $(\eta\in\R)$ can be represented
by the following series, that converges in the sense of the $L^2$--norm:
\beq
\label{Itrentotto}
\tJ^{(f)}_*(i\eta,\cdot) = \sum_{\ell=0}^\infty d_\ell\psi_\ell(\eta),
\eeq
where $\psi_\ell(\eta)$ are the Pollaczek functions defined by formula {\rm (\ref{Idieci})}, 
and the coefficients $d_\ell$ are given by:
\beq
\label{Itrentanove}
d_\ell=2\sqrt\pi\sum_{m=0}^\infty\frac{(-1)^m}{m!}\, 
\tcG^{(+,f)}_{(2\pi)}\left(m+\frac{3}{2},\cdot\right)
P_\ell\left[-i\left(m+\frac{1}{2}\right)\right],
\eeq
$P_\ell$ denoting the Pollaczek polynomials.

ii{\rm)} The function $J^{(f)}_*(v,\cdot)$ can be represented by the following expansion 
that converges in the sense of $L^2$--norm:
\beq
\label{Iquaranta}
J^{(f)}_*(v,\cdot)=\sum_{\ell=0}^\infty b_\ell\Phi_\ell(v),~~~~~(v\in\R^+),
\eeq
where the coefficients $b_\ell$ are given by $b_\ell=d_\ell/\sqrt{2\pi}$, and the 
functions $\Phi_\ell(v)$ are defined by formula {\rm (\ref{Iventidue})}.

iii{\rm)} The function $J^{(+,f)}_{(0)}(v,\cdot)$ can be represented by the following expansion:
\beq
\label{Iquarantuno}
J^{(+,f)}_{(0)}(v,\cdot) = e^v \sum_{\ell=0}^\infty b_\ell\Phi_\ell(v),~~~~~(v\in\R^+),
\eeq
that converges in the sense of the $L^2_{(w)}[0,+\infty)$--norm with weight function 
$w(v)=e^{-2v},\, (v\in\R^+)$.
\end{theorem}

\begin{proof}
The proof runs exactly as in the case of the boson statistics, with the only
remarkable difference that we use the Froissart--Gribov equalities (\ref{Itrentasette})
instead of (\ref{Iotto}).
\qed
\end{proof}

We can reconstruct, by the use of this method, the function $\tJ^{(f)}_*(i\eta,\cdot)$
but not the function $\tJ^{(+,f)}_{(0)}(i\eta,\cdot)$, which is much more interesting 
from the physical viewpoint. In order to recover the function
$\tJ^{(+,f)}_{(0)}(i\eta,\cdot)$ we must introduce a more restrictive assumption, requiring
the function $\tJ^{(+,f)}_{(0)}(\zeta,\cdot)=\int_0^{+\infty}J_{(0)}^{(+,f)}(v,\cdot)e^{-\zeta v}\,dv$
to be holomorphic in the half--plane $\Real\zeta >-\gamma$ ($\gamma>0$). 
Accordingly, in place of the temperateness condition (\ref{dodici}) we assume that
$J_{(0)}^{(+,f)}(v,\cdot)$ belongs to $L^1[0,+\infty)\cap L^2[0,+\infty)$.
Here, for the sake of simplicity, we treat only the case of fermions; analogous 
considerations hold true also in the case of the boson statistics. 
We can thus suppose that the singularities of $\tJ^{(+,f)}_{(0)}(\zeta,\cdot)$, 
corresponding to the excited states, all lie in the half--plane $\Real\zeta < -\gamma$, 
$\gamma$ being the smallest damping factor of the spectrum (see Refs.~\cite{Abrikosov,Fetter}). 
If this is the case, $\tJ^{(+,f)}_{(0)}(i\eta,\cdot)$ is analytic, and,
moreover, belongs also to $L^2(-\infty,+\infty)$. We can thus state the following result.

\begin{theorem}
\label{newtheorem3}
Let us assume that $\tJ^{(+,f)}_{(0)}(\zeta,\cdot)$ is a function holomorphic in 
the half--plane $\Real\zeta>-\gamma$ $(\gamma>0)$; then $\tJ^{(+,f)}_{(0)}(i\eta,\cdot)$ 
can be represented by the following expansion that converges in the sense 
of the $L^2$--norm:
\beq
\label{nn2}
\tJ^{(+,f)}_{(0)}(i\eta,\cdot)=\sum_{\ell=0}^\infty d'_\ell \psi_\ell(\eta),
\eeq
where $\psi_\ell(\eta)$ are the Pollaczek functions defined by formula 
{\rm(}\ref{Idieci}{\rm)}, and the coefficients $d'_\ell$ are given by:
\beq
\label{nn3}
d'_\ell=2\sqrt\pi\sum_{m=0}^\infty\frac{(-1)^m}{m!}\,
\tcG^{(+,f)}_{(2\pi)}\left(m+\frac{1}{2},\cdot\right)
P_\ell\left[-i\left(m+\frac{1}{2}\right)\right],
\eeq
$P_\ell$ denoting the Pollaczek polynomials.
\end{theorem}

\begin{proof}
The proof is strictly analogous to the one followed for proving equality (\ref{Inove}), 
and successively adapted to the fermion statistics in order to obtain expansion 
(\ref{Itrentotto}). The only remarkable difference is that now in the expression of 
the coefficients $d'_\ell$ we have the terms
$\tcG^{(+,f)}_{(2\pi)}\left(m+\frac{1}{2},\cdot\right)$ instead of 
$\tcG^{(+,f)}_{(2\pi)}\left(m+\frac{3}{2},\cdot\right)$; therefore all the coefficients 
corresponding to $m=0,1,2,\ldots$, are involved in the determination of the function 
$\tJ^{(+,f)}_{(0)}(i\eta,\cdot)$.
\qed
\end{proof}

Analogous methods and results can be worked out for the function
$\tJ^{(-,f)}_{(0)}(i\eta,\cdot)$, assuming that $\tJ^{(-,f)}_{(0)}(\zeta,\cdot)$ 
is holomorphic in the half--plane $\Real\zeta<\gamma$ ($\gamma>0$). We are then 
able to reconstruct the difference 
$\tJ^{(+,f)}_{(0)}(i\eta,\cdot)-\tJ^{(-,f)}_{(0)}(i\eta,\cdot)$ which leads to 
the determination of the ``spectral density''~\cite{Yukalov}.

\section{Reconstruction of the Jump Function in Terms of a Finite Number of Fourier Coefficients}
\label{se:reconstruction}
Up to now we have assumed that all the Fourier coefficients are known, and, in addition, 
that they are noiseless; but this assumption is clearly unrealistic. We now suppose that 
only a finite number of coefficients are known within a certain degree of approximation. 
We focus our attention on the case of the boson statistics, and specifically on the 
results contained in Lemmas~\ref{lemma1} and~\ref{lemma2}, and Theorem~\ref{theorem2}. 
The case of the fermion statistics can be treated similarly. We can simplify the 
notation, without ambiguity, by putting: $\tcG^{(+,b)}_{(2\pi)}(m+1,\cdot) = g_m$,
$e^{v/2} J^{(b)}_*(v,\cdot) = F_*(v)$, and $J^{(+,b)}_{(0)}(v,\cdot)=F(v)$. Then, 
we denote by $g^{(\epsilon)}_m$ the Fourier coefficients $\tcG^{(+,b)}_{(2\pi)}(m+1,\cdot)$ 
when they are perturbed by noise. We now assume that only $(N+1)$ Fourier coefficients 
are known within an approximation error of order $\epsilon$: 
i.e. $|g^{(\epsilon)}_m-g_m|\leq\epsilon$ ($m=0,1,2,\ldots,N$).

We consider the following finite sums:
\beq
\label{IIuno}
a^{(\epsilon,N)}_\ell = 
\sqrt 2\sum_{m=0}^N\frac{(-1)}{m!}\,g^{(\epsilon)}_m P_\ell\left[-i\left(m+\frac{1}{2}\right)\right].
\eeq
Accordingly, we have $a^{(0,\infty)}_\ell = a_\ell$ (see (\ref{Iventuno})). We can then
prove the following lemma.

\begin{lemma}
\label{lemma3}
The following statements hold true:
\begin{eqnarray}
\label{IIdue}
\mbox{i{\rm)}} & \hspace{3.9cm} & \sum_{\ell=0}^\infty \left|a^{(0,\infty)}_\ell\right|^2 = 
\| F_* \|^2_{L^2[0,\infty)} = C, ~~~~~(C = \mbox{constant}).  \hspace{1.8truecm}\\
\label{IItre}
\mbox{ii{\rm)}} & ~~~ & \sum_{\ell=0}^\infty \left|a^{(\epsilon,N)}_\ell\right|^2 = +\infty. 
\hspace{1truecm}\\
\label{IIquattro}
\mbox{iii{\rm)}} & ~~~ & 
\lim_{\staccrel{\scriptstyle N\rightarrow \infty}{\scriptstyle \epsilon\rightarrow 0}}
a^{(\epsilon,N)}_\ell = a^{(0,\infty)}_\ell = a_\ell,~~~~(\ell=0,1,2,\ldots). 
\hspace{1truecm}
\end{eqnarray}
\begin{itemize}
\item[iv{\rm)}] If $k_0(\epsilon,N)$ is defined as
\beq
\label{IIcinque}
k_0(\epsilon,N) = 
\max \left\{k\in\N\, : \, \sum_{\ell=0}^k |a^{(\epsilon,N)}_\ell|^2 \leq C\right\},
\eeq
i.e. it is the largest integer such that
 $\sum_{\ell=0}^k \left|a^{(\epsilon,N)}_\ell\right|^2 \leq C$, then
\beq
\label{IIsei}
\lim_{\staccrel{\scriptstyle N\rightarrow \infty}{\scriptstyle \epsilon\rightarrow 0}} 
k_0(\epsilon,N) = +\infty.
\eeq
\item[v{\rm)}] The sum
\beq
\label{IIsette}
M_k^{(\epsilon,N)} = \sum_{\ell=0}^k \left|a^{(\epsilon,N)}_\ell\right|^2,~~~~~(k\in\N),
\eeq
satisfies the following properties:
\begin{itemize}
\item[a{\rm)}] it increases for increasing values of $k$;
\item[b{\rm)}] the following relationships hold true:
\end{itemize}
\beq
\label{IIotto}
M_k^{(\epsilon,N)} \geq \left|a^{(\epsilon,N)}_k\right|^2 \staccrel{\sim}{k\rightarrow\infty}
\frac{1}{(N!)^2}\,(2k)^{2N},~~~~(N \,\mbox{fixed}\,).
\eeq
\end{itemize}
\end{lemma}

\begin{proof}
(i) Equality (\ref{IIdue}) follows from the Parseval theorem applied to expansion
(\ref{Iventi}), and recalling that $F_*(v)$ belongs to $L^2(-\infty,+\infty)$. \\
(ii) Let us rewrite the sums $a_\ell^{(\epsilon,N)}$ as follows:
\beq
\label{IInove}
a_\ell^{(\epsilon,N)}=\sum_{m=0}^N b_m^{(\epsilon)} P_\ell\left[-i\left(m+\frac{1}{2}\right)\right],
\eeq
where $b_m^{(\epsilon)}=\sqrt{2}(-1)^m g_m^{(\epsilon)}/m!$. 
Now, we can write the following inequality:
\begin{eqnarray}
\label{IIdieci}
\left | a_\ell^{(\epsilon,N)} \right | & = &
\left | \sum_{m=0}^N b_m^{(\epsilon)} 
P_\ell\left[-i\left(m+\frac{1}{2}\right)\right]\right| \nonumber \\
& \geq & \left | b_N^{(\epsilon)} P_\ell\left[-i\left(N+\frac{1}{2}\right)\right]\right|
\cdot \left|1-\frac{\left|\sum_{m=0}^{N-1}b_m^{(\epsilon)}
P_\ell\left[-i\left(m+\frac{1}{2}\right)\right]\right|}
{\left|b_N^{(\epsilon)}P_\ell\left[-i\left(N+\frac{1}{2}\right)\right]\right|}\right|.
\end{eqnarray}
Let us now recall that in the Appendix of Ref.~\cite{DeMicheli} the asymptotic 
behaviour of the Pollaczek polynomials $P_\ell[-i(m+1/2)]$ for large values of $l$ 
(at fixed $m$) is proved to be:
\beq
\label{IIundici}
P_\ell\left[-i\left(m+\frac{1}{2}\right)\right] \staccrel{\sim}{\ell\rightarrow\infty}
\frac{(-1)^\ell i^\ell}{m!} \, (2\ell)^m.
\eeq
Therefore, we have:
\beq
\label{IIdodici}
\frac{\left|\sum_{m=0}^{N-1}b_m^{(\epsilon)}
P_\ell\left[-i\left(m+\frac{1}{2}\right)\right]\right|}
{\left|b_N^{(\epsilon)}P_\ell\left[-i\left(N+\frac{1}{2}\right)\right]\right|}
\leq\frac{\sum_{m=0}^{N-1}\left|b_m^{(\epsilon)}
P_\ell\left[-i\left(m+\frac{1}{2}\right)\right]\right|}
{\left|b_N^{(\epsilon)}P_\ell\left[-i\left(N+\frac{1}{2}\right)\right]\right|}
\staccrel{\sim}{\ell\rightarrow\infty} 
\sum_{m=0}^{N-1}\left|\frac{b_m^{(\epsilon)}}{b_N^{(\epsilon)}}\right|\frac{N!}{m!}(2\ell)^{m-N}
\staccrel{\longrightarrow}{\ell\rightarrow\infty} 0.
\eeq
From (\ref{IIdieci}), (\ref{IIundici}) and (\ref{IIdodici}) it follows that for $\ell$
sufficiently large:
\beq
\label{IItredici}
\left|a_\ell^{(\epsilon,N)}\right|
\staccrel{\sim}{\ell\rightarrow\infty}\frac{\left|b_N^{(\epsilon)}\right|}{N!} \, (2\ell)^N.
\eeq
Therefore, $\lim_{\ell\rightarrow\infty}\left|a_\ell^{(\epsilon,N)}\right|=+\infty$, 
and statement (ii) follows. \\
(iii) We can write the difference $a_\ell^{(0,\infty)}-a_\ell^{(\epsilon,N)}$ as follows:
\begin{eqnarray}
\label{IIquattordici}
a_\ell^{(0,\infty)}-a_\ell^{(\epsilon,N)}&=& \sqrt{2}\left\{
\sum_{m=0}^N\frac{(-1)^m}{m!}\, (g_m-g_m^{(\epsilon)})
P_\ell\left[-i\left(m+\frac{1}{2}\right)\right]\right . \nonumber \\
&+& \left .  \sum_{m=N+1}^\infty \frac{(-1)^m}{m!} g_m 
P_\ell\left[-i\left(m+\frac{1}{2}\right)\right]\right\}.
\end{eqnarray}
In view of the fact that the series 
$\sqrt{2} \sum_{m=0}^\infty \frac{(-1)^m}{m!} g_m P_\ell[-i(m+\frac{1}{2})]$
converges to $a_\ell^{(0,\infty)}$, it follows that the second term in 
bracket (\ref{IIquattordici}) tends to zero as $N\rightarrow \infty$. 
Concerning the first term, we may write the inequality:
\beq
\label{IIquindici}
\left|\sum_{m=0}^N\frac{(-1)^m}{m!}(g_m-g_m^{(\epsilon)})
P_\ell\left[-i\left(m+\frac{1}{2}\right)\right]\right|
\leq\epsilon\sum_{m=0}^N\frac{1}{m!}\left|
P_\ell\left[-i\left(m+\frac{1}{2}\right)\right]\right|,
\eeq
where the inequalities $\left|g_m-g_m^{(\epsilon)}\right|\leq\epsilon$, 
$(m=0,1,2,\ldots,N)$ have been used.
Next, by rewriting the Pollaczek polynomials $P_\ell[-i(m+1/2)]$ as
\beq
\label{IIsedici}
P_\ell\left[-i\left(m+\frac{1}{2}\right)\right] = 
\sum_{j=0}^\ell p_j^{(\ell)}\left(m+\frac{1}{2}\right)^j,
\eeq
and, substituting this expression in inequality (\ref{IIquindici}), we obtain:
\beq
\label{IIdiciassette}
\epsilon\sum_{m=0}^N\frac{1}{m!}\left[\sum_{j=0}^l\left|p_j^{(\ell)}\right|
\left(m+\frac{1}{2}\right)^j\right].
\eeq
Next, we perform the limit for $N\rightarrow\infty$. In view of the fact that
$\sum_{j=0}^l p_j^{(\ell)}(m+1/2)^j$ is finite, and the series 
$\sum_{m=0}^\infty (m+1/2)^j/m!$ converges, we can exchange the order of the sums and write:
\beq
\label{IIdiciotto}
\epsilon\sum_{j=0}^l\left|p_j^{(\ell)}\right|\sum_{m=0}^\infty\frac{1}{m!}
\left(m+\frac{1}{2}\right)^j.
\eeq
Finally, performing the limit for $\epsilon\rightarrow 0$, and recalling equality 
(\ref{IIquattordici}), statement (iii) is obtained. \\
(iv) From definition (\ref{IIcinque}) it follows, for $k_1=k_0+1$, that:
\beq
\label{IIdiciannove}
\sum_{\ell=0}^{k_1}\left| a_\ell^{(\epsilon,N)}\right|^2 > C.
\eeq
Statement (iv) (formula (\ref{IIsei})) is proved if we can show that
$\lim_{\stackrel{{\scriptstyle N\rightarrow\infty}}{\epsilon\rightarrow 0}} k_1(\epsilon,N)=+\infty$.
Let us suppose that 
$\lim_{\stackrel{{\scriptstyle N\rightarrow\infty}}{\epsilon\rightarrow 0}} k_1(\epsilon,N)$ 
is finite. Then there should exist a finite number $K$ (independent of $\epsilon$ and $N$) such that,
for $N$ tending to $\infty$ and $\epsilon$ tending to zero, $k_1(\epsilon,N)\leq K$.
Then, from inequality (\ref{IIdiciannove}) we have:
\beq
\label{IIventi}
C < \sum_{\ell=0}^{k_1(\epsilon,N)}\left|a_\ell^{(\epsilon,N)}\right|^2\leq
\sum_{\ell=0}^{K}\left|a_\ell^{(\epsilon,N)}\right|^2.
\eeq
But as $N\rightarrow\infty, \epsilon\rightarrow 0$ we have (recalling also statement
(iii) formula (\ref{IIquattro})):
\beq
\label{IIventuno}
C < \sum_{\ell=0}^{K}\left|a_\ell^{(0,\infty)}\right|^2\leq
\sum_{\ell=0}^{\infty}\left|a_\ell^{(0,\infty)}\right|^2=C,
\eeq
which leads to a contradiction. Then statement (iv) follows. \\
(v) Concerning statement (a), it follows obviously from definition (\ref{IIsette}) of
$M_k^{(\epsilon,N)}$.
Finally, the first relationship in (\ref{IIotto}) is obvious; the second one follows
from the asymptotic behavior of $P_\ell[-i(m+1/2)]$ at large $\ell$ (for fixed $m$), 
i.e. formula (\ref{IIundici}).
\qed
\end{proof}

\begin{remark}
From statement (v) and formula (\ref{IIsei}) it follows that the sum $M_k^{(\epsilon, N)}$
presents, for large values of $N$ and small values of $\epsilon$, a plateau for $k \sim k_0$.
\end{remark}

By truncating expansion (\ref{Iventi}) we may now introduce an approximation of the function
$F_*(v)$ of the following type:
\beq
\label{IIventidue}
F_*^{(\epsilon,N)}(v)=\sum_{\ell=0}^{k_0(\epsilon,N)} a_\ell^{(\epsilon,N)} \Phi_\ell(v),
~~~~~(v\in\R^+).
\eeq
Approximation $F_*^{(\epsilon,N)}(v)$ is defined through the truncation number 
$k_0(\epsilon,N)$; the latter can be numerically determined by plotting the sum 
$M_k^{(\epsilon,N)}$ versus $k$, and exploiting properties (a) and (b), proved 
in statement (v) of the previous lemma and the property stated in the remark above 
(see also Ref.~\cite{DeMicheli}).

Now, we want to prove that the approximation $F_*^{(\epsilon,N)}(v)$ converges 
asymptotically to $F_*(v)$ in the sense of the $L^2$--norm, as $N\rightarrow\infty$ 
and $\epsilon\rightarrow 0$. We can prove the following theorem.

\begin{theorem} 
\label{theorem4}
The equality
\beq
\label{IIventitre}
\lim_{\staccrel{N\rightarrow\infty}{\scriptstyle\epsilon\rightarrow 0}}
\left\|F_* - F_*^{(\epsilon,N)}(v)\right\|_{L^2[0,+\infty)} = 0
\eeq
holds true.
\end{theorem}

\begin{proof}
From the Parseval equality it follows that:
\beq
\label{IIventiquattro}
\left\|F_*-F_*^{(\epsilon,N)}\right\|_{L^2[0,+\infty)}^2 =
\left\{\sum_{\ell=k_0+1}^{\infty}\left|a_\ell^{(0,\infty)}\right|^2 +
\sum_{\ell=0}^{k_0}\left|a_\ell^{(\epsilon,N)}-a_\ell^{(0,\infty)}\right|^2\right\}.
\eeq
Since $\sum_{\ell=0}^\infty\left|a_\ell^{(0,\infty)}\right|^2=C$ and
$\lim_{\stackrel{{\scriptstyle N\rightarrow\infty}}{\epsilon\rightarrow 0}} k_0(\epsilon,N)=+\infty$,
it follows that
$\lim_{\stackrel{{\scriptstyle N\rightarrow\infty}}{\epsilon\rightarrow 0}}
\sum_{\ell=k_0+1}^{\infty} \left|a_\ell^{(\epsilon,N)}\right|^2 = 0.$
It is convenient to rewrite the second term of the r.h.s. of (\ref{IIventiquattro})
as follows. Let us define:
\beq
\label{IIventicinque}
h_\ell^{(0,\infty)} = \left\{
\begin{array}{ll}
a_\ell^{(0,\infty)} & \mbox{if $\ell$ is even,} \\
-i a_\ell^{(0,\infty)} & \mbox{if $\ell$ is odd,}
\end{array}
\right . 
\eeq
\beq
\label{IIventisei}
h_\ell^{(\epsilon,N)} = \left\{
\begin{array}{ll}
a_\ell^{(\epsilon,N)} & \mbox{if $\ell$ is even,} \\
-i a_\ell^{(\epsilon,N)} & \mbox{if $\ell$ is odd.}
\end{array}
\right . 
\eeq
Notice that $h_\ell^{(0,\infty)}$ and $h_\ell^{(\epsilon,N)}$ are real, and
$\sum_{\ell=0}^{k_0}\left |a_\ell^{(\epsilon,N)}-a_\ell^{(0,\infty)}\right |^2 =
\sum_{\ell=0}^{k_0}\left (h_\ell^{(\epsilon,N)}-h_\ell^{(0,\infty)}\right )^2$. Next, we 
introduce the following functions:
\begin{eqnarray}
H^{(0,\infty)}(v)=\sum_{\ell=0}^{\infty}h_\ell^{(0,\infty)} 
\UU_{[\ell,\ell+1[}(v), \label{IIventisettea} \\
H^{(\epsilon,N)}(v)=\sum_{\ell=0}^{\infty}h_\ell^{(\epsilon,N)} 
\UU_{[\ell,\ell+1[}(v), \label{IIventisetteb}
\end{eqnarray}
where $\UU_E$ is the characteristic function of the set $E$. From statements (i),
(ii) and (iii) of the previous lemma (formulae (\ref{IIdue}), (\ref{IItre}) and
(\ref{IIquattro})) we obtain:
\beq
\label{IIventotto}
\int_0^{+\infty} \left (H^{(0,\infty)}(v)\right )^2\, dv = 
\sum_{\ell=0}^\infty \left (h_\ell^{(0,\infty)}\right )^2 = C,
\eeq
\beq
\label{IIventinove}
\int_0^{+\infty} \left (H^{(\epsilon,N)}(v)\right )^2\, dv = 
\sum_{\ell=0}^\infty \left (h_\ell^{(\epsilon,N)}\right )^2 = +\infty,
\eeq
\beq
\label{IItrenta}
H^{(\epsilon,N)}(v) \staccrel{\longrightarrow}
{\stackrel{{\scriptstyle N\rightarrow\infty}}{\epsilon\rightarrow 0}}
H^{(0,\infty)}(v),~~~~~ (v\in [0,+\infty)).
\eeq
Hereafter, we assume, for the sake of simplicity and without loss of
generality, that every term $h_\ell^{(\epsilon,N)}$ is different from zero. Next, let
$V(\epsilon,N)$ be the unique root of equation
$\int_0^V\left (H^{(\epsilon,N)}(v)\right)^2\,dv=C$. Let us indeed observe that
$\int_0^V\left (H^{(\epsilon,N)}(v)\right)^2\,dv$ is a continuous non--decreasing 
function which is zero for $V=0$, and $+\infty$ for $V\rightarrow +\infty$. 
Furthermore, from statement (iv) of the previous lemma (formula (\ref{IIsei})) we have 
$\lim_{\stackrel{{\scriptstyle N\rightarrow\infty}}{\epsilon\rightarrow 0}} V(\epsilon,N)=+\infty$. \\
Then we can write:
\begin{eqnarray}
\label{IItrentuno}
\int_{0}^{V(\epsilon,N)}\left[H^{(\epsilon,N)}(v)-H^{(0,\infty)}(v)\right]^2\,dv = 
\hspace{6.5cm} \nonumber \\ 
=\int_{V(\epsilon,N)}^{+\infty}\left(H^{(0,\infty)}(v)\right)^2dx
-2\int_{0}^{V(\epsilon,N)} H^{(0,\infty)}(v)
\left[H^{(\epsilon,N)}(v)-H^{(0,\infty)}(v)\right ]\, dv.
\end{eqnarray}
Next, we perform the limit for $N\rightarrow\infty$ and $\epsilon\rightarrow 0$.
Concerning the first term at the r.h.s. of (\ref{IItrentuno}) we have:
\beq
\label{IItrentadue}
\lim_{\stackrel{{\scriptstyle N\rightarrow\infty}}{\epsilon\rightarrow 0}}
\int_{V(\epsilon,N)}^{+\infty} \left (H^{(0,\infty)}(v)\right )^2\,dv = 0.
\eeq
For what concerns the second term, we introduce the following function:
\beq
\label{IItrentatre}
B^{(\epsilon,N)}(v)= \left\{
\begin{array}{ll}
H^{(\epsilon,N)}(v)-H^{(0,\infty)}(v) & ~~~\mbox{if $0\leq v\leq V(\epsilon,N)$}, \\
0 & ~~~\mbox{if $v > V(\epsilon,N)$}.
\end{array}
\right .
\eeq
Then, we have by the use of the Schwarz inequality
\beq
\label{IItrentaquattro}
\int_{0}^{+\infty} \left | B^{(\epsilon,N)}(v)\right |^2\, dv \leq 4C,
~~~~~~ (N < \infty, \epsilon > 0).
\eeq
Moreover, from (\ref{IItrenta}) we have:
\beq
\label{IItrentacinque}
B^{(\epsilon,N)}(v) 
\staccrel{\longrightarrow}{\stackrel{{\scriptstyle N\rightarrow\infty}}{\epsilon\rightarrow 0}} 0~,
~~~~~ v \in [0,+\infty).
\eeq
The family of functions $\{B^{(\epsilon,N)}(v)\}$ is bounded in $L^2[0,+\infty)$, 
therefore it has a subsequence which is weakly convergent in $L^2[0,+\infty)$.
The limit of this subsequence is zero. 
In fact, let us observe that $|B^{(\epsilon,N)}(v)| \leq 2C$; then we consider the 
function $B^{(\epsilon,N)}(v)\phi(v)$, where $\phi$ is an arbitrary element of the 
class of functions $C_c^\infty(\R^+)$. We then have 
$|B^{(\epsilon,N)}(v)\phi(v)|\leq 2C|\phi(v)|$, and this inequality does not
depend on $N$ and $\epsilon$. In view of the Lebesgue dominated convergence theorem 
we can then write (see also limit (\ref{IItrentacinque})):
\beq
\label{nn8bis}
\lim_{\stackrel{{\scriptstyle N\rightarrow\infty}}{\epsilon\rightarrow 0}}\sup\left|
\int_0^{+\infty}B^{(\epsilon,N)}(v)\phi(v)\,dv\right|=0.
\eeq
Since the set of functions $C_c^\infty(\R^+)$ is everywhere dense in $L^2[0,+\infty)$, 
given an arbitrary function $\psi\in L^2[0,+\infty)$ and an arbitrary number $\eta>0$, 
there exists a function $\phi_k\in C_c^\infty(\R^+)$ such that 
$\left\|\psi-\phi_k\right\|_{L^2[0,+\infty)} < \eta$.
Furthermore, through the Schwarz inequality we have:
\begin{eqnarray}
\label{nn9}
\int_0^{+\infty}\left| B^{(\epsilon,N)}(v)[\phi_k(v)-\psi(v)]\right| \,dv & \leq &
\left(\int_0^{+\infty}\left|B^{(\epsilon,N)}(v)\right|^2\,dv\right)^{1/2}
\left(\int_0^{+\infty}\left|\phi_k(v)-\psi(v)\right|^2\,dv\right)^{1/2} \nonumber \\
& \leq & 2\sqrt{C}\,\eta.
\end{eqnarray}
From (\ref{nn8bis}) and (\ref{nn9}) we can conclude that
\beq
\label{nn10}
\lim_{\stackrel{{\scriptstyle N\rightarrow\infty}}{\epsilon\rightarrow 0}}\sup\left|
\int_0^{+\infty}B^{(\epsilon,N)}(v)\psi(v)\,dv\right|=0,
\eeq
for any $\psi\in L^2[0,+\infty)$.

Next, by using the same type of arguments, we can state that if there is an 
arbitrary subsequence belonging to the family $\{B^{(\epsilon,N)}\}$ that weakly 
converges in $L^2[0,+\infty)$, then the weak limit of this subsequence is 
necessarily zero. Finally, from the uniqueness of the (weak) limit point,
it follows that the whole family $\{B^{(\epsilon,N)}\}$ converges weakly to zero 
in $L^2[0,+\infty)$. 

We can thus write:
\beq
\label{nn11}
\lim_{\stackrel{{\scriptstyle N\rightarrow\infty}}{\epsilon\rightarrow 0}}
\int_0^{+\infty} H^{(0,\infty)}(v) B^{(\epsilon,N)}(v)\, dv = 0,
\eeq
and from equality (\ref{IItrentuno}) we have
\beq
\label{IItrentotto}
\lim_{\stackrel{{\scriptstyle N\rightarrow\infty}}{\epsilon\rightarrow 0}}
\int_0^{V(\epsilon,N)}\left[H^{(\epsilon,N)}(v)-H^{(0,\infty)}(v)\right]^2\, dv = 0.
\eeq
Since $\sum_{\ell=0}^{k_0}\left|a_\ell^{(\epsilon,N)}-a_\ell^{(0,\infty)}\right|^2
\leq\int_0^{V(\epsilon,N)}\left[H^{(\epsilon,N)}(v)-H^{(0,\infty)}(v)\right]^2\,dv$,
we have:
\beq
\label{IItrentanove}
\lim_{\stackrel{{\scriptstyle N\rightarrow\infty}}{\epsilon\rightarrow 0}}
\sum_{\ell=0}^{k_0}\left |a_\ell^{(\epsilon,N)}-a_\ell^{(0,\infty)}\right |^2 = 0,
\eeq
and, in view of equality (\ref{IIventiquattro}), the theorem is proved.
\qed
\end{proof}

We can then prove the following corollary.
\begin{corollary}
\label{corollary4}
The following equality holds true:
\beq
\label{IIquaranta}
\lim_{\stackrel{{\scriptstyle N\rightarrow\infty}}{\epsilon\rightarrow 0}}
\left\|F(v)-e^{v/2}\sum_{\ell=0}^{k_0(\epsilon,N)}a_\ell^{(\epsilon,N)}
\Phi_\ell(v)\right\|_{L^2_{(w)}[0,+\infty)}=0,
\eeq
$L^2_{(w)}[0,+\infty)$ being the weighted $L^2$--space with weight function 
$w(v)=e^{-v}$, $(v\in\R^+)$, and the functions $\Phi_\ell(v)$ are defined by 
formula {\rm (\ref{Iventidue})}.
\end{corollary}

\begin{proof}
The statement follows immediately from Theorem~\ref{theorem4} by noting that:
\begin{eqnarray}
\label{nn12}
\int_0^{+\infty}\left|F_*(v)-\sum_{\ell=0}^{k_0(\epsilon,N)}
a_\ell^{(\epsilon,N)}\Phi_\ell(v)\right|^2\,dv
&=& \int_0^{+\infty}e^{-v}\left|F(v)-e^{v/2}\sum_{\ell=0}^{k_0(\epsilon,N)}a_\ell^{(\epsilon,N)}
\Phi_\ell(v)\right|^2\,dv \nonumber \\
&=&\left\|F(v)-e^{v/2}\sum_{\ell=0}^{k_0(\epsilon,N)}a_\ell^{(\epsilon,N)}
\Phi_\ell(v)\right\|^2_{L^2_{(w)}[0,+\infty)}.
\end{eqnarray}
We can thus conclude that the jump function
$J_{(0)}^{(+,b)}(v,\cdot)=F(v)$ can be approximated by the truncated expansion
\beq
\label{nn13}
J_{(0)}^{(+,b)}(v,\cdot) \sim 
e^{v/2}\sum_{\ell=0}^{k_0(\epsilon,N)}a_\ell^{(\epsilon,N)}\Phi_\ell(v),~~~~~~(v\in\R^+). 
\hspace{1cm} \Box
\eeq
\end{proof}

%
%
%
%
%
%
%
%
%

\newpage

\begin{thebibliography}{32}

\bibitem{Abrikosov}
Abrikosov, A.A., Gorkov, L.P. and Dryaloshinski, I.E.: 
{\it Methods of Quantum Field Theory in Statistical Physics.}
Englewood Cliffs: Prentice--Hall, 1963

\bibitem{Bateman}
{\it Bateman Manuscript Project: Higher Trascendental Functions.} A. Erdelyi, Director. Vol. \textbf{2}.
New York: Krieger, 1953

\bibitem{Boas}
Boas, R.P.:
{\it Entire Functions.}
New York: Academic Press, 1954

\bibitem{Bremermann}
Bremermann, H.:
{\it Distributions, Complex Variables, and Fourier Transforms.}
Reading: Addison--Wesley, 1965

\bibitem{Bros1}
Bros, J. and Buchholz, D.:
Axiomatic Analyticity Properties and Representations of Particles in Thermal Quantum Field Theory.
Ann. Inst. H. Poincar\'e -- Physique Theorique {\bf 64}, 495--521 (1996)

\bibitem{Bros2}
Bros, J. and Viano, G.A.:
Connection Between the Harmonic Analysis on the Sphere and the Harmonic Analysis on 
the One--sheeted Hyperboloid: An Analytic Continuation Viewpoint.
I Forum Mathematicum {\bf 8}, 621--658 (1996),
II Forum Mathematicum {\bf 8}, 659--722 (1996),
III Forum Mathematicum {\bf 9}, 165--191 (1997)

\bibitem{Bros3}
Bros, J. and Buchholz, D.:
Fields at finite temperature: A general theory of the two--point functions.
In preparation

\bibitem{Buchholz}
Buchholz, D. and Junglas, P:
On the Existence of Equilibrium States in Local Quantum Field Theory.
Commun. Math. Phys. {\bf 121}, 255--270 (1989)

\bibitem{DeMicheli}
De Micheli, E. and Viano, G.A.:
On the Solution of a Class of Cauchy Integral Equations.
J. Math. Anal. Appl. {\bf 246}, 520--543 (2000)

\bibitem{Fetter}
Fetter, A.L. and Walecka, J.D.:
{\it Quantum Theory of Many--Particle Systems.}
New York: McGraw--Hill, 1971

\bibitem{Haag1}
Haag, R., Hugenholtz, N.M. and Winnink, M.:
On the Equilibrium States in Quantum Statistical Mechanics.
Commun. Math. Phys. {\bf 5}, 215--236 (1967)

\bibitem{Hoffman}
Hoffman, K.:
{\it Banach Spaces of Analytic Functions.}
Englewood Cliffs: Prentice--Hall, 1962

\bibitem{Itzykson}
Itzykson, C.:
Group Representation in a Continuous Basis: An Example.
J. Math. Phys. {\bf 10}, 1109--1114 (1969)

\bibitem{LeBellac}
Le Bellac, M.:
{\it Thermal Field Theory.}
Cambridge: Cambridge Univ. Press, 1996

\bibitem{Matsubara}
Matsubara, T.:
A new approach to quantum--statistical mechanics.
Prog. Theor. Phys. {\bf 14}, 351--378 (1955)

\bibitem{Szego}
Szeg\"{o}, G.:
{\it Orthogonal Polynomials.}
New York: Academic Press, 1954

\bibitem{Yukalov}
Yukalov, V.I.:
{\it Statistical Green's Functions.}
Kingston: Queen's University Press, 1998

\end{thebibliography}
\end{document}